\documentclass[submission,copyright,creativecommons]{eptcs}
\providecommand{\event}{POST 2017 -- 6th International Conference on Principles of Security and Trust} 
\pdfoutput=1
\usepackage{amsmath}
\usepackage{amssymb}
\usepackage{amsfonts}
\usepackage{color}

\usepackage[margin=20mm,top=20mm,bottom=30mm]{geometry} 

\usepackage{url}
\usepackage{graphicx}
\usepackage{wrapfig}
\usepackage{hyperref}

\newcommand{\enc}{\texttt{enc}}

\newcommand{\senc}{\texttt{senc}} 
\newcommand{\sdec}{\texttt{sdec}}  
\newcommand{\pub}{\texttt{pub}}

\newcommand{\h}{\texttt{h}}
\renewcommand{\tt}[1]{\texttt{#1}}
\newcommand{\nonce}{\texttt{nonce}}
\newcommand{\N}{{\cal N}}

\newcommand{\send}[2]{\texttt{out}(#1,#2)}
\newcommand{\recv}[2]{\texttt{in}(#1,#2)}

\newcommand{\F}{\mathcal{F}}
\newcommand{\M}{\mathcal{M}}
\newcommand{\C}{\mathcal{C}}
\renewcommand{\P}{\mathcal{P}}

\newtheorem{theorem}{Theorem}[section]
\newtheorem{lemma}[theorem]{Lemma}
\newtheorem{proposition}[theorem]{Proposition}
\newtheorem{corollary}[theorem]{Corollary}

\newcommand{\Letin}[2]{\texttt{let\; } #1=#2 \texttt{\; in\; }} 
\newcommand{\pilet}{\texttt{let\;}}

\newcommand{\pin}{\texttt{\;in\;}}

\newcommand{\pvnew}[1]{\texttt{new}\,#1}   

\newcommand{\Ifthen}[2]{\texttt{if } #1=#2 \texttt{ then }}
\newcommand{\Else}{\texttt{\;else\;}} 

\newcommand{\piIfthen}[2]{\texttt{if } #1=#2 \texttt{ then }} 
\newcommand{\piIfthenp}[1]{\texttt{if } #1 \texttt{ then }} 
 
\newcommand{\piIfthenelsep}[3]{\piIfthenp{#1} #2 \Else #3}

\newcommand{\E}{\mathcal{E}}

\newcommand{\rto}{\rightarrow}

\newcommand{\seal}{\texttt{seal}}

\newcommand{\sig}{\texttt{sig}}

\newcommand{\state}{\texttt{state}}

\newcommand{\smram}{\texttt{smram}}
\newcommand{\getPCR}{\texttt{pcr}}
\newcommand{\getINT}{\texttt{int}}
\newcommand{\getCACHE}{\texttt{cache}}
\newcommand{\getINIT}{\texttt{init}}
\newcommand{\getPP}{\texttt{pp}}

\newcommand{\getLOCK}{\texttt{lock}}
\newcommand{\getSMIH}{\texttt{smi}}
\newcommand{\getSTM}{\texttt{stm}}
\newcommand{\isSMALL}{\texttt{is\_small}}
\newcommand{\isBIG}{\texttt{is\_big}}

\newcommand{\setPCR}{\texttt{set\_pcr}}
\newcommand{\setINT}{\texttt{set\_int}}
\newcommand{\setINIT}{\texttt{set\_init}}
\newcommand{\setPP}{\texttt{set\_pp}}
\newcommand{\setLOCK}{\texttt{set\_lock}}

\newcommand{\tpmAccess}{\texttt{tpm\_acc}}
\newcommand{\tpmch}{\texttt{tpm\_ch}}
\newcommand{\cputpm}{\texttt{cpu\_tpm}}

\newcommand{\cpuAccess}{\texttt{cpu\_acc}}

\newcommand{\cache}{\texttt{cache}}
\newcommand{\flush}{\texttt{flush}}

\newcommand{\TC}{\texttt{drt}}

\newcommand{\stm}{\texttt{stm}}
\newcommand{\smih}{\texttt{smi}}

\newcommand{\init}{\texttt{init}}
\newcommand{\pp}{\texttt{pp}}

\newcommand{\true}{\texttt{true}}
\newcommand{\false}{\texttt{false}}

\newcommand{\tpm}{\texttt{tpm}}

\newcommand{\cpu}{\texttt{cpu}}
\newcommand{\unseal}{\texttt{unseal}}

\newcommand{\priv}{\texttt{priv}}

\newcommand{\resetPCR}{\texttt{reset}}
\newcommand{\extendPCR}{\texttt{extend}}

\newcommand{\data}{\texttt{data}}

\newcommand{\hello}{\texttt{hi}}

\newcommand{\Att}{\texttt{Att}}

\renewcommand{\top}{\texttt{top}}

\newcommand{\prog}{\texttt{prog}}
\newcommand{\program}{\texttt{prog}}

\newcommand{\getentry}{\texttt{get\_entry}}

\newcommand{\pfstate}{\texttt{pf\_state}}
\newcommand{\Comp}{\texttt{Comp}}
\newcommand{\drt}{\texttt{drt}}
\newcommand{\ps}{\tt{p}_\tt{s}}
\newcommand{\pd}{\tt{p}_\tt{d}}
\renewcommand{\cache}{\texttt{cache}}
\newcommand{\length}{\texttt{length}}

\newcommand{\fdata}{\texttt{data}}
\newcommand{\chain}{\texttt{chain}}
\newcommand{\Tinit}{\texttt{Tinit}}
\newcommand{\Tpp}{\texttt{Tpp}}
\newcommand{\Tstm}{\texttt{Tstm}}

\renewcommand{\int}{\texttt{int}}

\newcommand{\drtpp}{\texttt{PP}
}
\newcommand{\drtinit}{\texttt{INIT}
}

\newcommand{\CPU}{\texttt{CPU}}

\newcommand{\INT}{\texttt{INT}}
\newcommand{\LOCK}{\texttt{LOCK}}
\newcommand{\PCR}{\texttt{PCR}}
\newcommand{\RESET}{\texttt{RESET}}
\newcommand{\EXTEND}{\texttt{EXTEND}}

\newcommand{\PP}{\texttt{PP}}
\newcommand{\INIT}{\texttt{INIT}}
\newcommand{\SETUP}{\texttt{SETUP}}
\newcommand{\DATA}{\texttt{DATA}}
\newcommand{\EXEC}{\texttt{EXEC}}
\newcommand{\DRT}{\texttt{DRT}}
\newcommand{\CACHE}{\texttt{CACHE}}
\newcommand{\TPM}{\texttt{TPM}}
\newcommand{\UNSEAL}{\texttt{UNSEAL}}
\newcommand{\STM}{\texttt{STM}}
\newcommand{\SMI}{\texttt{SMI}}
\newcommand{\SMIH}{\texttt{SMIH}}
\newcommand{\SMRAM}{\texttt{SMRAM}}
\newcommand{\SMALL}{\texttt{SMALL}}
\newcommand{\BIG}{\texttt{BIG}}
\renewcommand{\sec}{\texttt{sec}}

\newcommand{\todo}[1]{} 
\newcommand{\longVersion}[1]{#1}

     \newcommand{\cjd}[1]{}
        \renewcommand{\sb}[1]{} 

\begin{document}
      


 \title{Automated verification of dynamic root of trust protocols \\ (long version)} 

\author{Sergiu Bursuc
\institute{University of Bristol, UK}
\and Christian Johansen 
\institute{Dept. of Informatics, University of Oslo}
\and Shiwei Xu 
\institute{Wuhan Digital Engineering Institute, China}
}
\def\titlerunning{Automated verification of dynamic root of trust protocols}
\def\authorrunning{
S.~Bursuc
\& C.~Johansen
\& S.~Xu
}

\maketitle

\begin{abstract}

Automated verification of security protocols based on dynamic root of trust, typically relying on protected hardware such as TPM, involves several challenges that we address in this paper. We model the semantics of trusted computing platforms (including CPU, TPM, OS, and other essential components) and of associated protocols in a classical process calculus accepted by ProVerif. As part of the formalization effort, we introduce new equational theories for representing TPM specific platform states and dynamically loaded programs. 

%
Formal models for such an extensive set of features cannot be readily handled by ProVerif, due especially to the search space generated by unbounded extensions of TPM registers. In this context we introduce a transformation of the TPM process, that simplifies the structure of the search space for automated verification, while preserving the security properties of interest. This allows to run ProVerif on our proposed models, so we can 
derive automatically security guarantees for protocols running in a dynamic root of trust context.

%
\end{abstract}

\tableofcontents

\todo{cite warinschi and kremer in related work} 

\section{Introduction}

A hardware root of trust, including dynamic measurement of programs and their protected execution, is a promising concept for ensuring the integrity of a platform and the privacy of sensitive data, despite powerful software attackers \cite{tc}. This relies on the idea that hardware is more difficult to compromise than software, and therefore, it can play a crucial role in protocols for handling sensitive data. When a secure computing platform is needed, a special sequence of instructions allows for a trusted piece of hardware to attest the integrity of the software to be run and to give access to data in a protected environment.  

However, turning this idea into a secure design and implementation is not easy, as various attacks have shown \cite{txt-att1,txt-att2}. For more assurance, one could use models and tools that allow automated verification of desired properties against trusted computing protocols and implementations. 
One main challenge for automated verification is the size and number of components involved in running programs protected by a dynamic root of trust. 
Furthermore, messages of such protocols consist not only of data, but also of programs that are to be executed on the platform, and that can be supplied by an attacker or by an honest participant. 
At the same time, modelling the platform configuration registers (\PCR) of the trusted platform module (\TPM) \cite{tc-tpm} poses problems, because $\PCR$s can be extended an unbounded number of times. Even the most efficient symbolic methods struggle with the structure of the resulting search space \cite{proverif,dkrs-tpm}.  

%

\todo{summary for our models of state and dynamic loading of programs, and relate it to other models of state propose by Ryan and Kremer}

\emph{Our contributions.} We propose a formal model in the ProVerif process calculus \cite{blanchet-corr} for the technology and for the security properties of a dynamic root of trust (as instantiated by Intel's Trusted Execution Technology or AMD's Secure Virtual Machine). Our model is more realistic than \cite{dkrs-tpm} and it covers aspects of trusted computing that \cite{datta-tc} does not cover (section \ref{sec:model}). We show 
how a platform state can be naturally represented as a term in ProVerif (or applied pi-calculus \cite{abadi01popl,ryan-smyth-pi}) and how operations on the platform state can be expressed as equations in a term algebra (sections \ref{sec:state-model} and \ref{sec:state-access}). Furthermore, we show 
how to model the dynamic loading of protected programs. 
Our model is simple and does not require heavy encodings, being based on the classic idea of processes as data, with a twist to take protection into account (section \ref{sec:prog-model}). 


\sb{intro for abstraction needs to be revised, most likely it is not an abstraction, but a sound and complete transformation of processes; also we need better reference to \cite{dkrs-tpm}}
We propose a new abstraction to model the extension of \PCR\ registers 
that allows automated verification for a larger class of protocols than in \cite{dkrs-tpm}. We show how to over-approximate the model of the \TPM\ such that the structure of the search space is simplified, without losing possible attacks or introducing false attacks. The main idea is that we can let the attacker set the PCR to \emph{any} value, as long as it is ``big enough'' (section \ref{sec:pcr-abstraction}). 

Putting the formalisation and the abstraction together, we obtain the first automated verification for a realistic model of a dynamic root of trust. As security properties, we prove code integrity (the PCR values correctly record the measurement of the platform) and secrecy of sealed data (only a designated program can access data that has been sealed for its use in a protected environment). 

\vspace{2ex}
\noindent\textbf{Acknowledgements: } 
We would like to thank Cas Cremers and several reviewers for helping improve this work.

\section{Related work}\label{sec:related}

%

A programming language and a logic for specifying trusted computing protocols and properties are proposed in \cite{datta-tc}. The setting is quite expressive and it allows the analysis of protocols similar to the ones that we study in this paper. \cite{datta-tc} does not consider the seal/unseal functions of the \TPM, but their language could be extended to capture them. However, the formal analysis of \cite{datta-tc} is manual, and considering the complexity of the proofs involved, the lack of automation can be a limitation. We also believe some of their axioms (like those linking the \PCR\ values to a late launch action) could be decomposed into more atomic formulas, in closer relation to the computational platform. Their security properties include correctly reading \PCR\ values and the ability of honest parties to launch roots of trust; our property of code integrity, modeled as a correspondence assertion, can be seen as an additional constraint for these two events. 

The analysis of \cite{dkrs-tpm} is automated with ProVerif and is based on a Horn clause model. Microsoft's Bitlocker protocol is shown to preserve the secrecy of data sealed against a static sequence of PCR values. 
Their model considers a static root of trust, and cannot handle dynamically loaded programs.
Furthermore, there is no way to express a program that has access to data in a protected environment. 
Without a richer model of platform states, code integrity properties cannot be expressed either. 
%
%
To help with automation, \cite{dkrs-tpm} shows that, for a specific class of Horn clauses, it is sound to bound the number of extensions of PCR registers. Since our model is in applied pi-calculus and our security properties are different, we cannot directly rely on their result, and we propose a new way of handling the unbounded \PCR\ extension problem. 


\emph{Information-flow security and computational models.} \cite{fournet-tpm} presents a secure compiler for translating programs and policies into cryptographic implementations, distributed on several machines equipped with TPMs. A computational model capturing functionalities similar to ours, in conjunction with additional features such as authenticated key exchange, was recently proposed in \cite{warinschi}. Our models are more abstract, yet could be related to particular implementations - a closer connections between formal and computational models could be explored in future. 

\emph{Unbounded search space.} Several works tackle the problem of an unbounded search space for automated verification, but technically they are all based on principles that cannot be translated to PCR registers. In \cite{miriam-lists}, it is shown that, for a class of Horn clauses, verification of protocols with unbounded lists can be reduced to verification of protocols with lists containing a single element. In \cite{cortier-topo}, it is shown that to analyse routing protocols it is sufficient to consider topologies with at most four nodes. These are strong results, based on the fact that the elements of a list or the nodes in a route are handled uniformly by the protocol. Similar results, in a different context, are shown in \cite{franklin-sp,franklin-post}. Their reductions are based on the principle of data independence for memory stores. In \cite{kusters-diffie} and respectively \cite{post-renc}, it is shown how to handle an unbounded number of Diffie-Hellman exponentiations and respectively reencryptions in ProVerif. Surprisingly, the underlying associative-commutative properties of Diffie-Hellman help in \cite{kusters-diffie}, while \cite{post-renc} can rely on the fact that a re-encryption does not change the semantics of a ciphertext. 
Another case where an unbounded number of operations is problematic is file sharing \cite{proverif-plutus}. In order to obtain an automated proof, \cite{proverif-plutus} assumes a bound on the number of access revocations, without providing justifications for soundness. A sound abstraction for an unbounded number of revocations, in a more general setting, is proposed in \cite{set-abs}. Still, it is specialized to databases and it seems to rely on the same principle as several results mentioned above: it does not matter what the data is, it only matters to what set it belongs. 

 
\emph{Tools and models for non-monotonic state.} StatVerif \cite{statverif} is aimed specifically for the verification of protocols relying on non-monotonic states, encoding the semantics of applied pi-calculus enriched with states into a set of Horn clauses for input to ProVerif. Tamarin \cite{SchmidtMCB12Tamarin} is based on multiset rewriting and inherently allows specification and automated reasoning for non-monotonic states, where the set of facts can both augment and decrease. SAPIC \cite{KremerK14GlobalState} takes as input a stateful variant of applied pi-calculus and produces a multiset-based model, which is then analysed using Tamarin.       

StatVerif \cite{statverif}, SAPIC \cite{KremerK14GlobalState}, and Tamarin directly \cite{meier}, have been used with success to verify security protocols that rely on non-monotonic states or trusted hardware: $PKCS\sharp11$ for key management  \cite{pkcs}, YubiKey for user authentication \cite{yubikey}, and protocols for contract signing \cite{contract}. Our models, on the other hand, are tailored for direct input to ProVerif, while extending the scope of formal models for platform state operations and dynamic root of trust protocols based on a TPM \cite{tc-book,tc,tc-tpm}. It is one of our main interests for future work to see how the models of this paper can be analysed with tools like \cite{SchmidtMCB12Tamarin,KremerK14GlobalState,statverif}, in order to obtain a closer alignment with the state semantics of real systems.

\section{Preliminaries}

\subsection{Trusted computing}\label{sec:prelim-drt} 

We first describe the required computing platform (hardware and software) and then describe the considered class of 
dynamic root of trust protocols.

\textbf{A. Computing platform.} We consider a general purpose computing platform equipped with a \CPU\ and a \TPM\ (both trusted), as well as a generic untrusted operating system.

\emph{Trusted hardware.}  Trusted computing relies on the \CPU\ and the \TPM\footnote{See recent book \cite{tpm20book} detailing the TPM version 2.0 specification and implementations.} to perform certain operations whose integrity cannot be compromised by any software attacker.
Regarding the \TPM, two of its trusted features are fundamental for the applications that we consider in this paper: the ability to record a chain of values in its \emph{platform configuration registers} (\PCR) and the ability to \emph{seal data} against specified values of the \PCR. 


The \TPM\ allows the \PCR\ to be \emph{reset} only by the \CPU\ or by a system reset. On the other hand, the \PCR\ can be \emph{extended} with any value by software. If a \PCR\ records a value $p$ and is extended with a value $v$, the new value of the \PCR\ is $\h((p,v))$, i.e. the result of applying a hash function to the concatenation of $p$ and $v$. Crucially, these are the only two ways in which the values of a \PCR\ can be modified. The role of the \PCR\ for the protocols that we consider in this paper is to store the measurement of programs, recording a chain of loaded programs. When data $d$ is \emph{sealed} against some specified value $v$ of the \PCR, the \TPM\ stores $d$ internally and can release it in future only if the value recorded in its \PCR\ matches the value $v$ against which $d$ was sealed.   

For the purpose of formal verification, we are flexible about who exactly of the \CPU\ or the \TPM\ is doing a trusted operation, like measuring, sealing, etc. This depends on the implementation, e.g., the Intel SGX can do all the operations of a \TPM. Changing the formalization from this paper to fit a particular implementation should be easy.

\emph{Privileged software.} 
When a system interrupt is triggered (e.g by network communication or user interface action), all physical memory can be accessed by the system management interrupt (\SMI) handler. This means that any memory protection mechanism, in particular the protocols that we consider in this paper, must either disable interrupts for their whole duration (not practical in general) or else rely on the fact that the \SMI\ handler cannot be compromised. That is why the \SMI\ handler is stored in a memory area called \SMRAM, which enjoys special hardware protection. Still, as shown in \cite{txt-att1,txt-att2}, the security guarantees of trusted computing can be violated using the \CPU\ caching mechanism to compromise the \SMI\ handler. Roughly, these attacks work because the protection of the \SMRAM\ is not carried on to its cached contents. 
A countermeasure against such attacks, that we also adopt in this paper at an abstract level, is a software transfer monitor (\STM) \cite{tc-book}. It also resides in the \SMRAM, but it cannot be cached while a dynamic root of trust is running (special registers of the \CPU\ should ensure that), and its role is to protect some memory regions from the \SMI\ handler.

\begin{figure}[t]
\begin{center}
\centerline{\hspace{-0.5ex}
   \resizebox{0.6\textwidth}{!}{
  \includegraphics{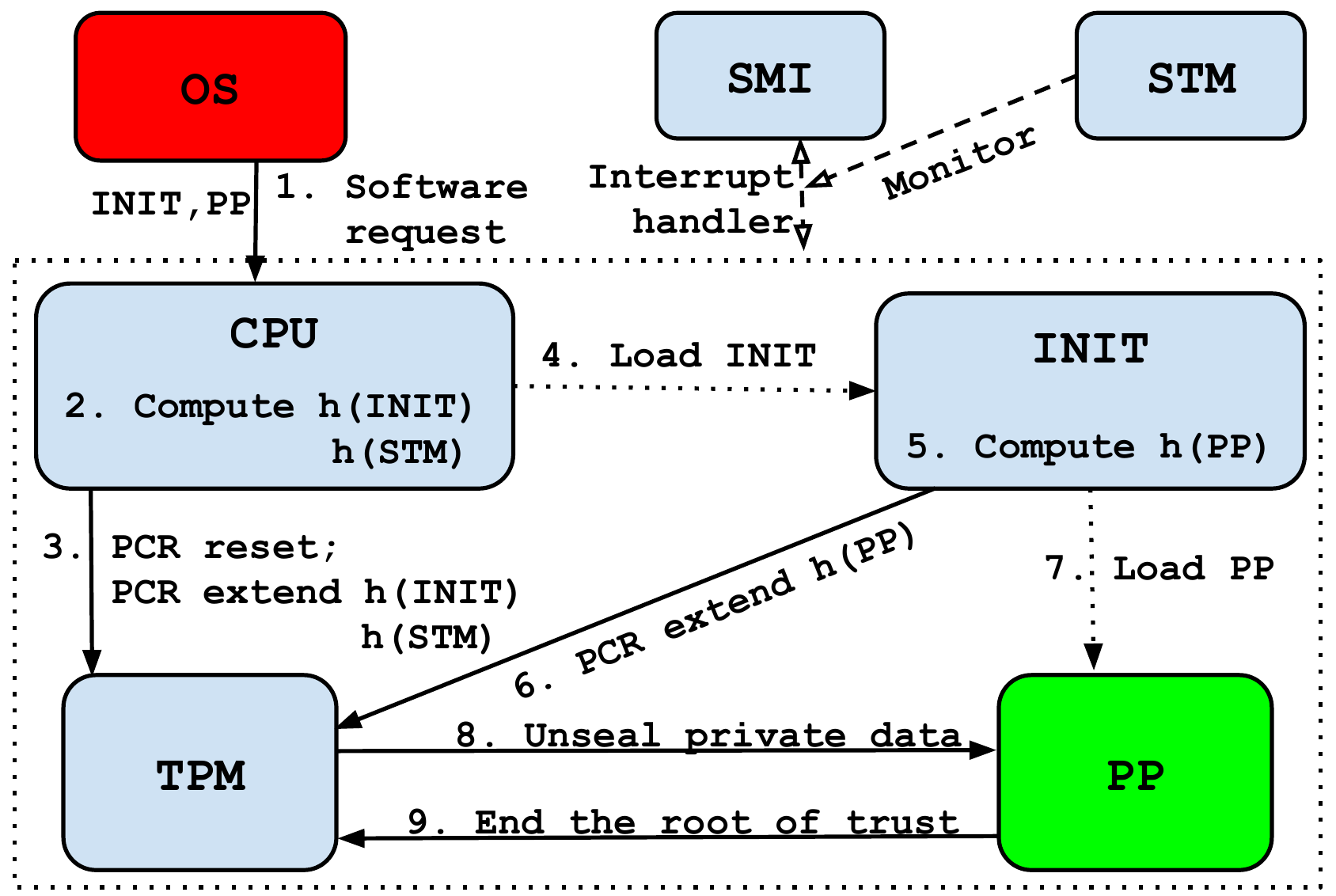}}\hspace{0.5ex}%
\begin{tabular}{|c|c|c|} 
\multicolumn{3}{c}{\vspace{-3cm}}\\
\multicolumn{3}{c}{\textbf{Trust assumptions}}\\
\hline
\textbf{Untrusted} & \textbf{Measured} & \textbf{Trusted} \\
\hline 
\texttt{OS} & \texttt{INIT} & \texttt{CPU} \\
\hline
\texttt{SMI} & \texttt{PP} & \texttt{TPM}\\
\hline
& \texttt{STM} &\\
\hline
\end{tabular}
}
\end{center}
\vspace{-0.5cm}
\caption{Execution flow in a Dynamic Root of Trust (DRT)\label{fig:drt}}
\end{figure}

\textbf{B. Dynamic root of trust.} We consider the technology of dynamic measurement and protected execution, also called dynamic root of trust (\DRT), as instantiated for example in Intel's \tt{Trusted Execution Technology (TXT)} or AMD \tt{Secure Virtual Machine (SVM)}, and as illustrated in Fig.~\ref{fig:drt}.  

The goal of \DRT\ is to establish a protected execution environment for a program, where private data can be accessed without being leaked to an attacker that controls the operating system. Assume a program, that we will call \drtpp\ (called \tt{measured launch environment} on Intel and \tt{secure kernel} on \tt{AMD}), needs to be loaded in a protected environment. The first entry point of the \DRT\ protocol is a trusted instruction of the \CPU\ (called \tt{GETSEC[SENTER]} on Intel and \tt{SKINIT} on \tt{AMD}), that takes as input the program \drtpp. 
To help with the establishment of a protected environment, the \CPU\ also receives as input another program, that we will call \drtinit\ (called \tt{SINIT authenticated code module} on Intel and \tt{secure loader} on \tt{AMD}). The \DRT\ launch and execution sequence can then be summarized as follows:

\vspace{-2ex}
\begin{description}
\item[1.] The \CPU\ receives a request from the operating system containing the \drtinit\ code and the \drtpp\ code. The system interrupts are disabled at this step, as an additional protection against untrusted interrupt handlers.

\item[2-3.] A software attacker that controls the operating system could compromise \drtinit\ and the \STM, and that is why the \CPU\ computes their measurement and extends the result into the \TPM, to keep a trace of programs responsible for the \DRT. Measuring a program means applying a hash function to its source code. This computation is performed on the \CPU\ and is trusted, 
 entailing that the resulting value is a correct measurement of \drtinit\ and \STM. The \CPU\ communicates with the \TPM\ on a trusted channel and requests that the \PCR\ is reset and extended with the resulting value (\h(\drtinit),\h(\STM)).   

\item[4-7.] The \drtinit\ program is loaded and it computes the measurement of the \drtpp\ program, extending it into the \PCR. The communication between \drtinit\ and the \TPM\ is performed on a private channel established by the \CPU. \drtinit\ also allocates protected memory for the execution of  \drtpp\ and loads it.  

\item[8.] The \drtpp\ program can re-enable interrupts once appropriate interrupt handlers are set. Furthermore, it can now request the \TPM\ to unseal data that has been sealed against the current \PCR\ value, and it can have access to that data in a protected environment. The communication between \drtpp\ and the \TPM\ is performed on a private channel established by the \CPU. 

\item[9.] Before ending its execution, the \drtpp\ program extends the \PCR\ with a dummy value, to record that the platform state is not to be trusted any more.  
\end{description}

Since the OS is untrusted it can supply malicious programs \drtinit\ and \drtpp. Therefore, \drtinit, \drtpp\ and the \STM\ are not trusted, but they are \emph{measured}. If their measurement does not correspond to some expected \emph{trusted} values, this will be recorded in the \TPM\ and secret data will not be unsealed for this environment. 

\textbf{Security goals.} Let us summarize the two main security goals of the \DRT. 


\emph{Code integrity:} In any execution of the platform, if the measurements recorded in the \PCR\ value of the \TPM\ correspond to the sequence of programs $\P_\drtinit$, $\P_\STM$, $\P_\drtpp$, then the platform is indeed running a \DRT\ for the protected execution of $\P_\drtpp$ in the context of $\P_\drtinit$ and $\P_\STM$. In particular, this means that the programs $\P_\drtpp$, $\P_\drtinit$ and $\P_\STM$ cannot be modified while a \DRT\ is running.

\emph{Secrecy of sealed data:} Any secret data that is sealed only against a \PCR\ value recording the sequence of programs $\P_\drtinit$, $\P_\STM$, $\P_\drtpp$, is only available for the program $\P_\drtpp$, in any execution of the platform. 

%


\subsection{ProVerif process calculus}

We review ProVerif \cite{proverif,blanchet-corr} and the special way in which we use (a restriction of) its input calculus in our modelling. 

\textbf{A. Terms, equational theories and deducibility.} We consider an infinite set of \emph{names}, $a, b, c, k, n \ldots$, an infinite set of \emph{variables}, $x, y, z, \ldots$ and a
possibly infinite set of \emph{function symbols} $\F$. Names and variables are \emph{terms}; new terms are built by applying
function symbols to names, variables and other terms. 
We split $\F$ into two disjoint sets of \emph{public} functions $\F^\pub$ and \emph{private} functions $\F^\priv$. Public functions can be applied by anyone to construct terms, including the attacker, whereas private functions can be applied only as specified by the protocol. When $\F^\priv$ is not explicit, we assume that all functions are public.

A \emph{substitution} $\sigma$ is a partial function from variables to terms. The replacement of every variable $x$ with $x\sigma$ in a term $T$ is denoted by $T\sigma$. A \emph{context} is a term $\C[\_]$ that contains a special symbol $\_$ in place of a subterm. For a context $\C[\_]$ and a term $T$, we denote by $\C[T]$ the term obtained by replacing $\_$ with $T$ in $\C[\_]$. For any formal object $\cal D$, we denote by $\sig({\cal D})$ the set of function symbols appearing in $\cal D$, and by $\top(T)$ the outer-most function symbol in term $T$.  


En equational theory $\E$ is defined by a set of rewrite rules $U_1\rto V_1,\ldots,U_n\rto V_n$, where $U_1,\ldots,U_n,V_1,\ldots,V_n$ are terms with variables. 
A term $U$ rewrites to $V$ in one step, denoted by $U\rto V$, if there is a context $\C[\_]$, a substitution $\sigma$ and an index $i\in \{1,\ldots,n\}$ such that $U=\C[U_i\sigma]$ and $V=C[V_i\sigma]$. Several rewrite steps from $U$ to $V$ are denoted by $U\rto^*V$. 
We consider only convergent equational theories, i.e., for any term $T$ there exists a unique non-reducible term $T\!\!\downarrow$ s.t.\ $T\rto^{*} T\!\!\downarrow$. We write $U=_\E V$ iff $U\!\!\downarrow = V\!\!\downarrow$.
ProVerif also allows operations on sequences: for all $n$, from any terms $T_1,\ldots,T_n$, one can derive the term $(T_1,\ldots,T_n)$, and conversely. 

\emph{Deduction.} Given an equational theory $\E$, a set of terms $S$ and a term $T$, the ability of an attacker to obtain $T$ from $S$ is captured by the deduction relation $S\vdash_\E T$ (or simply $S\vdash T$ when $\E$ is understood) defined as being true iff: \\
\noindent $\bullet$ there exists a term $T'\in S$ such that $T'=_\E T$, or \\
\noindent $\bullet$ there are terms $T_1,\ldots,T_n$ such that $S\vdash_\E T_1,\ldots,S\vdash_\E T_n$ and a function symbol $f\in \F^\pub$ such that ${f(T_1,\ldots,T_n)=_\E T}$


%
%
%

\vspace{1ex}
\textbf{B. Processes and operational semantics.}\label{sec:processes} 

\begin{figure}[b]
\begin{center}
\begin{tabbing}
$P, Q, R ::= $ \; \;  \\           
\begin{tabular}{rcl}
   \begin{tabular}{ll}
    $0$ \; & null process \\
   $P \mid Q$ \; & parallel composition \\
   $!P$ \; & replication \\
   $\pvnew{n};P$ \; & name restriction \\
  \end{tabular} &&
\begin{tabular}{ll}
  \; $\recv{U}{T};P$ \; & message input on $U$ \\
  \; $\send{U}{T};P$ \; & message output on $U$ \\
  \; $\piIfthenelsep{U = V}{P}Q$ \; & conditional \\
  \; $\Letin{x}{T}{P}$ \; & term evaluation \\
\end{tabular}
\end{tabular}
\end{tabbing}
\end{center}
\caption{Process algebra, with $n$ a name, $x$ a variable, and $T,U,V$ terms. \label{fig:processes}}
\end{figure}

\emph{Processes} of the calculus are built according to Fig.~\ref{fig:processes}.
Replication spawns instances of a process: $!P$ is formally equivalent with $P \;|\; !P$. Names introduced by $\pvnew{}$ are called \emph{bound} or \emph{private}; they represent the creation of fresh data. Names that are not bound are called \emph{free}, or \emph{public}. 
The term $T$ in an input $\recv{U}{T}$ allows to specify filters for messages received on $U$: a message $M$ will be accepted only if there is a substitution $\sigma$ such that $M=T\sigma$. A variable $x$ is \emph{free} in a process $P$ if $P$ neither contains $x$ in any of its input patterns nor does it contain any term evaluation of the form $x=T$. Consecutive term evaluations can be written together as $\Letin{(x_1,\ldots,x_n)}{(T_1,\ldots,T_n)}{P}$. The notions of substitution, contexts and normal forms translate to processes as expected. 

\emph{Operational semantics} is defined as a transition system on configurations of the form $(\N,\M,\P)$, where: $\N$ is a set of fresh names created during the execution of a process;
$\M$ is the set of terms made available to the attacker; and $\P$ is the set of processes executing in parallel at a given point in time. We write
 $(\N,\M,\P)\rto^*(\N',\M',\P')$ if the configuration $(\N',\M',\P')$ can be reached from $(\N,\M,\P)$ in zero or more executions steps. 
Such a sequence of execution steps is called a trace of $P$.

\textbf{C. Security properties.} The ability of an attacker to learn a term $T$ by interacting with a process $P$ is denoted by $P\models \Att(T)$, defined as true iff there exists a process $Q$, with $\sig(Q)\cap\F^\priv=\emptyset$, such that 
$(\N_{\tt{init}},\emptyset,\{P\mid Q\})\rto^* (\N',\M',\P')$ and $\M\vdash_\E T$, for some configuration $(\N',\M',\P')$. Intuitively, $Q$ represents any computation that can be performed by the attacker. 

A (simplified) \emph{correspondence assertion} \cite{blanchet-corr} is a formula of the form 
\[\Att(T)\implies false \;\;\;\;\text{or}\;\;\;\;\Att(T)\implies (U=V).\]
For a correspondence assertion $\Att(T)\implies \Phi$ as above, we have 
\[P\models \Att(T)\implies \Phi \;\;\;\;\text{iff}\;\;\;\;\forall \sigma.\;\;[\;\; (P\models \Att(T\sigma)) \implies \Phi\sigma\;\;]\] 

Correspondence assertions of the first type model the \emph{secrecy} of $T$, while those of second type enforce the constraint $U=V$ for deducible terms matching the pattern $T$ (typically the terms $U,V$ will share variables with $T$). 

\section{Formalisation}\label{sec:model}

Our formal specification for the trusted computing platform and protocols described in section \ref{sec:prelim-drt} assumes an attacker that controls the operating system and can execute a \DRT\ any number of times, with any \drtinit\ and \drtpp\ programs. Moreover, using the \CPU\ cache, the attacker can compromise the \STM\ and \SMI\ handler, and use them 
to access protected memory. The attacker has access to all \TPM\ functions. 
However, we assume that the attacker cannot compromise the \CPU\ nor the \TPM,
and that the platform state can only be modified according to the equations that we present in section \ref{sec:state-access}.

We model a system state as a term that can be updated by the \CPU\ process, the \TPM\ process and, once it has been output on a public channel, by the attacker. Multiple system states can be explored in parallel by the attacker, whose knowledge monotonically accumulates the set of all reachable states. This is an abstraction with respect to a real platform, where the \CPU\ and the \TPM\ have their own internal state, part of a global, non-monotonic system state.  We also have a simplified model of \TPM\ sealing: in reality, it relies on encryption with a TPM private key and refers to a specific system state; in our model, it is represented by the pair of public/private functions $\seal$/$\unseal$. For unsealing, the \TPM\ process will require the input of a system state and check that the corresponding unseal request is valid for that state.





\subsection{Cryptographic primitives and platform constants}


To model cryptographic primitives and various constants on the platform state, we consider the signature $\F_\data$, where $\F_\fdata^\priv = \{\unseal/2\}$ and 
\[\F_\fdata^\pub=\{\ps/0,\pd/0,\true/0,\false/0,\h/1,\senc/2,\sdec/2,\seal/2\}.\]
We also consider the set of rewrite rules $\E_\fdata$:  
\[
\begin{array}{rcl}
\sdec(\senc(x_{\tt{val}},x_{\tt{key}}),x_\tt{key}) & \rto & x_\tt{val}\\
\unseal(\seal(x_{\tt{val}},x_{\tt{pcr}}), x_{\tt{pcr}}) & \rto  & x_{\tt{val}} 
\end{array}
\]
The constant $\pd$ (resp. $\ps$) represents the result of a dynamic (resp. static) \PCR\ reset. A dynamic reset marks the start of a dynamic root of trust, and can only be performed by the \CPU. The functions $\senc$ and $\sdec$, and the corresponding rewrite rule, model symmetric key encryption. 
The symbol $\h$ represents a hash function. 
Anyone can seal a value, while the corresponding rewrite rule and the fact that $\unseal$ is private ensure that a value can be unsealed only according to the specification of the \TPM. 

\subsection{Dynamically loaded programs}\label{sec:prog-model}

%
%
%

To model the fact that arbitrary programs can be dynamically loaded on the platform state (e.g. for the roles of $\INIT$ and $\PP$), we consider a new public function symbol $\prog/1$ and an infinite signature of \emph{private constants $\F_\P$}, containing a different constant $n_P$ for every possible process $P$. Intuitively, the term $\prog(n_P)$ is a public and unique identifier for the program $P$. In a computational model, such an identifier can for example be obtained by hashing the source code of $P$. The first action of a process that models a program will be to output the corresponding program identity $\prog(n_P)$ on a public channel.


On the other hand, the constant $n_P$ represents a \emph{private entry point} for the program $P$. 
Specifically, we consider a private function $\getentry$ and the rewrite rule $\getentry(\prog(x))\rto x$. The idea is that a trusted loader of programs (the \CPU\ in our case) has access to the private function $\getentry$ and, using this rewrite rule, it can gain access to the private entry point of any program. Now, $n_P$ can play the role of a private channel between the trusted loader and the loaded program. 
Furthermore, 
we can store program identifiers in the platform state, to record what programs are loaded. Then, we can rely on $n_P$ to model the ability of certain loaded programs to affect the platform state (shown in section \ref{sec:state-access}). We denote by $\E_\prog$ the equational theory defined in this subsection: \hspace{1ex} $\F_\prog=\{\prog/1\}\cup \F_\P\ , \hspace{3ex} \E_\prog=\{\getentry(\prog(x))\rto x\}$.

\subsection{Platform state}\label{sec:state-model} 


To model a platform state, we consider the signature:
\[
\begin{array}{rcl}
\F_{\state} & = & \{\state/4,\tpm/1,\cpu/2,\smram/2,\drt/3\} 
\end{array}
\]
where all the symbols of $\F_{\state}$ are private. This ensures that a platform state can be constructed or modified only according to the specification, relying on equations that we present in subsection \ref{sec:state-access}.
Intuitively, a term of the form 
\[
\begin{array}{rl}
\state(&\tpm(T_\PCR),\cpu(T_\INT,T_\CACHE),\smram(T_\STM,T_\SMIH),\drt(T_\INIT,T_\PP,T_\LOCK))
\end{array}
\]
represents a platform state where:\\
\noindent $\bullet$ \textbf{$T_\PCR$} is a term that represents the value of the \PCR\ register of the \TPM;\\
\noindent $\bullet$ \textbf{$T_\INT$} is the value of a register of the \CPU\ showing if interrupts are enabled; \\
\noindent $\bullet$ \textbf{$T_\CACHE$} represents the contents of the \CPU\ cache; \\
\noindent $\bullet$ \textbf{$T_\SMIH$} represents the program for the \SMI\ handler and \textbf{$\STM$} represents the \STM\ program, which are located in \SMRAM;\\
\noindent $\bullet$ \textbf{$T_\LOCK$} is showing if a dynamic root of trust is running; \\
\noindent $\bullet$ \textbf{$T_\INIT$} represents the \drtinit\ program; \\
\noindent $\bullet$ \textbf{$T_\PP$} represents the protected program \drtpp.



\subsection{Read and write access}\label{sec:state-access} The read access is universal: any agent who has access to a platform state 
\[
\state(\tpm(T_\PCR),\cpu(T_\INT,T_\CACHE),\smram(T_\STM,T_\SMIH),\drt(T_\INIT,T_\PP,T_\LOCK))
\]
can read any of its components relying on the public unary function symbols 
$$\F_{\tt{read}}=\{\getPCR, \getINT, \getCACHE,\getSTM,\getSMIH,\getINIT,\getPP,\getLOCK\}$$ 
and associated rewrite rules:

\[
\begin{array}{rcl} 
\getPCR(\state(\tpm(y),x_1,x_2,x_3)) & \rto & y \\ 
\getINT(\state(x_1,\cpu(y_1,y_2),x_2,x_3)) & \rto & y_1 \\
\getCACHE(\state(x_1,\cpu(y_1,y_2),x_2,x_3)) & \rto &  y_2 \\ 
\getINIT(\state(x_1,x_2,\drt(y_1,y_2,y_3),x_3)) & \rto & y_1 \\
\getPP(\state(x_1,x_2,\drt(y_1,y_2,y_3),x_3)) & \rto & y_2 \\
\getLOCK(\state(x_1,x_2,\drt(y_1,y_2,y_3),x_3)) & \rto & y_3 \\ 
\getSTM (\state(x_1,x_2,x_3,\smram(y_1,y_2))) & \rto & y_1 \\ 
\getSMIH (\state(x_1,x_2,x_3,\smram(y_1,y_2))) & \rto & y_2  
\end{array}
\]

\begin{figure}[!b]
\caption{Write access to the platform state \label{fig:eq-set}}
\[
\begin{array}{c}
\centerline{
   \resizebox{0.5\textwidth}{!}{
  \includegraphics{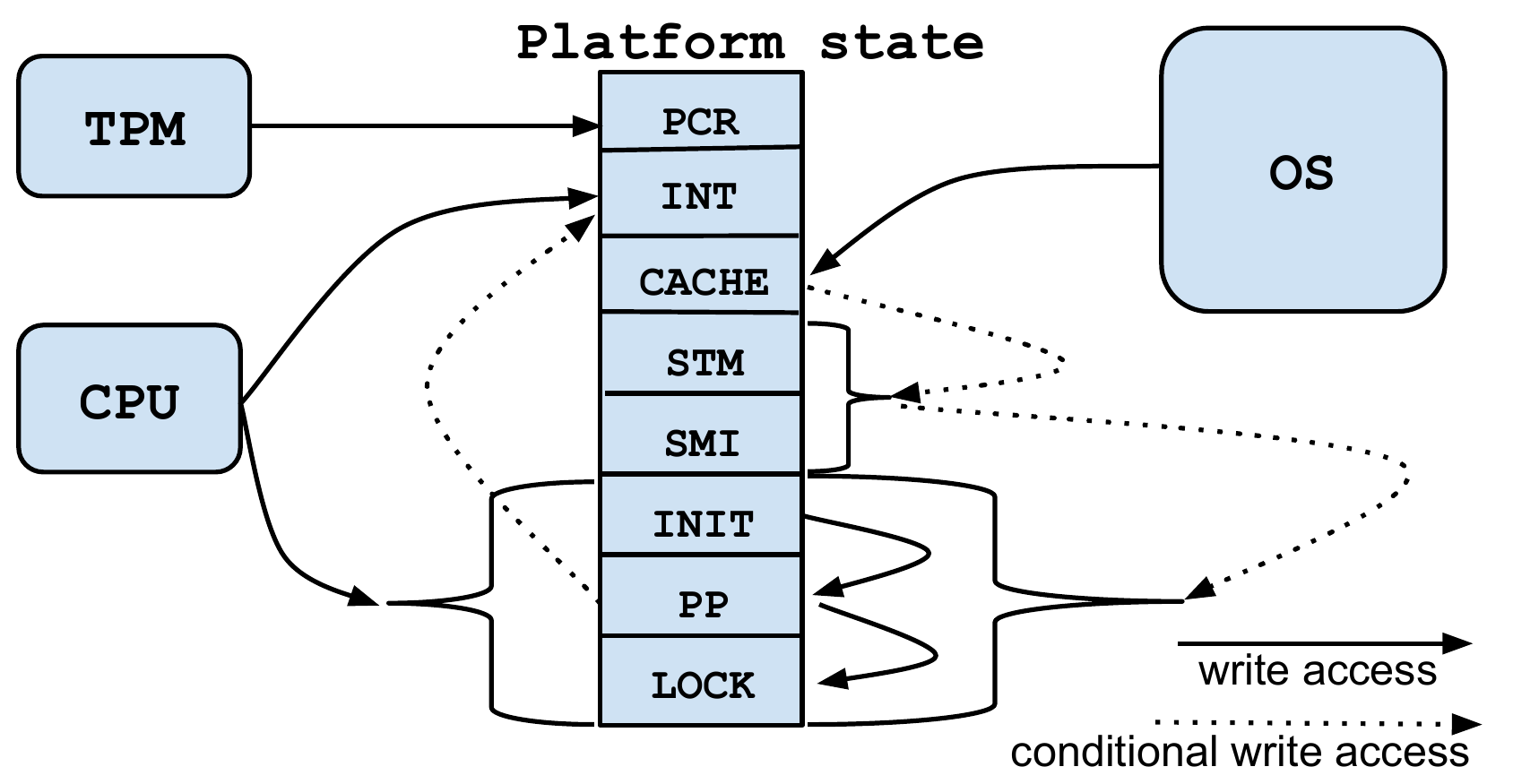}
    }}\\
\begin{array}{l} 
 \resetPCR(\state(\tpm(y),x_1,x_2,x_3),\tpmAccess,\ps) \rto \state(\tpm(\ps),x_1,x_2,x_3)\\
\resetPCR (\state(\tpm(y),x_1,x_2,x_3),\tpmAccess,\pd) \rto \state(\tpm(\pd),x_1,x_2,x_3)\\
\extendPCR(\state(\tpm(y),x_1,x_2,x_3), \tpmAccess, v) \rto \state(\tpm(\h((y,v))),x_1,x_2,x_3)\\
\setPCR(\state(\tpm(y),x_1,x_2,x_3), \tpmAccess, v) \rto \state( \tpm(v),x_1,x_2,x_3)\\
\setINT(\state(x_1,\cpu(y_1,y_2),x_2,x_3),\cpuAccess,v) \rto \state(x_1,\cpu(v,y_2),x_2,x_3)\\
\setINT(\state(x_1,\cpu(y,z),x_2,\drt(z_1,\program(z_2),\true)),z_2,v)\\ \ \ \ \ \rto
\state(x_1,\cpu(v,z),x_2,\drt(z_1,\program(z_2),\true))\\
\cache(\state(x_1,\cpu(y_1,y_2),x_2,x_3),v) \rto \state(x_1,\cpu(y_1,v),x_2,x_3)\\
\flush\_\stm(\state(x_1,\cpu(y_1,v),\smram(z_1,z_2),\drt(w_1,w_2,\false)))\\ \ \ \ \ \rto
\state(x_1,\cpu(y_1,v),\smram(v,z_2),\drt(w_1,w_2,\false)))\\
\flush\_\smih(\state(x_1,\cpu(y_1,v),\smram(z_1,z_2),x_2))\\ \ \ \ \ \rto
\state(x_1,\cpu(y_1,v),\smram(z_1,v),x_2)\\
\setINIT(\state(x_1,x_2,x_3, \drt(y_1,y_2,y_3)), \cpuAccess,v)\\ \ \ \ \ \rto
\state(x_1,x_2,x_3,\drt(v,y_2,y_3))\\
\setPP(\state(x_1,x_2,x_3,\drt(y_1,y_2,y_3)),\cpuAccess,v)\\ \ \ \ \ \rto
\state(x_1,x_2,x_3,\drt(y_1,v,y_3))\\
\setPP(\state(x_1,x_2,x_3,\drt(\program(y_1),y_2,y_3)),y_1,v)\\ \ \ \ \ \rto
\state(x_1,x_2,x_3,\drt(\program(y_1),v,y_3))\\
\setPP(\state(x,\cpu(\true,z),\smram(\program(z_1),\program(z_2))),\drt(y_1,y_2,y_3)),(z_1,z_2),v)\\ \ \ \ \ \rto
\state(x,\cpu(\true,z),\smram(\program(z_1),\program(z_2)),\drt(y_1,v,y_3))\\
\setLOCK(\state(x_1,x_2,x_3,\drt(y_1,y_2,y_3)),\cpuAccess,v)\\ \ \ \ \ \rto
\state(x_1,x_2,x_3, \drt(y_1,y_2,v))\\
\setLOCK(\state(x_1,x_2,x_3,\drt(y_1,\program(y_2),y_3)),y_2,v)\\ \ \ \ \ \rto
\state(x_1,x_2,x_3,\drt(y_1,\program(y_2),v))\\
\setLOCK(\state(x,\cpu(\true,z),\smram(\program(z_1),\program(z_2)),\drt(y_1,y_2,y_3)),(z_1,z_2),v)\\ \ \ \ \ \rto
\state(x,\cpu(\true,z),\smram(\program(z_1),\program(z_2)),\drt(y_1,y_2,v))
\end{array}
\end{array}
\]
\end{figure}

The write access to the platform state is restricted by the equational theory described and illustrated in Fig.~\ref{fig:eq-set}, where $\tpmAccess$ and $\cpuAccess$ are private constants and all other new symbols are public. 


\begin{description}
 \item[\PCR.] Only the \TPM\ can reset, extend or set the value of the \PCR. This capability of the \TPM\ is modeled by the \emph{private constant} \tpmAccess, which will be used only in the \TPM\ process, described later in Fig.~\ref{fig-TPM}.
 \item[\INT.] The interrupts can be enabled or disabled by the \CPU, whose capability is modeled by the \emph{private constant} \cpuAccess. Additionally, if a \DRT\ is running, then the corresponding protected program \drtpp\ also has the ability to enable or disable interrupts. This is modeled in the second $\setINT$ equation, by relying on the fact that, if $\prog(x)$ represents the public identity of a program (as explained in section \ref{sec:prog-model}), then $x$ represents a private entry point for that program. Therefore, we can use $x$ to model the ability of $\prog(x)$ to change certain elements of the platform state when it is loaded.
 \item[\CACHE.] Any values can be cached. The cache values can then be copied into the contents of the \SMI\ handler and, when a \DRT\ is not running, into the \STM\ component of the state.  
 \item[\INIT.] Only the \CPU\ has the ability to load an \INIT\ program on the platform.  
 \item[\PP.] The \PP\ program can be loaded by the \CPU\ (the first equation for \setPP) or by an \INIT\ program, if the latter is already loaded on the platform (the second equation for \setPP). Furthermore, the \SMI\ in conjunction with the \STM\ can also modify the \PP\ program, if the interrupts are enabled (the third equation for \setPP).      
 \item[\LOCK.] Similarly, the \DRT\ lock can be set/unset by the \CPU, by the running \PP, or by the \SMI\ in conjunction with the \STM, if the interrupts are enabled.  
\end{description}


  


We denote by $\E_{\state}$ the equational theory defined in this subsection. 

\subsection{Communication channels}  


The public constant $\tt{os}$ models a communication channel for platform states and other messages that may be intercepted, modified or provided by the intruder as inputs to the \CPU\ or the \TPM.  A private constant $\cputpm$ models the secure channel between the \CPU\ and the \TPM. A private function $\tpmch$ models the ability of the \CPU\ to establish a private channel between a loaded program and the \TPM. Generally, these channels will be of the form $\tpmch(\program(t))$ and the CPU will send this term both to the program represented by $\program(t)$ (on channel $t$) and to the TPM (on channel $\cputpm$). We also use message tags that will be clear from the context. 

\subsection{The trusted platform module}

\begin{figure}
\[
\begin{array}{|ll|} 
\hline
\TPM & =\;\; !\TPM_\RESET\;|\;!\TPM_\EXTEND \;|\; !\TPM_\UNSEAL\\
\TPM_\RESET &= \pilet (\mathit{ch}, \mathit{rv})\!=\!(\cputpm,\pd) \pin !\PCR_\RESET \; |\\
&\hspace{-0.5cm}\pilet (\mathit{ch}, \mathit{rv}) = (\tt{os},\ps) \pin !\PCR_\RESET\\

\PCR_\RESET & = \recv{\mathit{ch}}{(\texttt{reset\_req}, \mathit{nonce}, \mathit{pf\_state})};\\
&\hspace{-0.5cm}\pilet \mathit{new\_st} = \resetPCR(\mathit{pf\_state},\tpmAccess,\mathit{rv}) \pin  \\
&\hspace{-0.5cm}\send{\mathit{ch}}{(\texttt{reset\_resp}, \mathit{nonce}, \mathit{new\_st})}\\

\TPM_\EXTEND &= \pilet \mathit{ch} = \cputpm \pin !\PCR_\EXTEND\; |\\
&\hspace{-0.5cm}\pilet \mathit{ch} = \tt{os} \pin !\PCR_\EXTEND\; |\\
&\hspace{-0.5cm}!\; (\recv{\cputpm}{(\tt{ext\_channel}, \mathit{ch})}; !\PCR_\EXTEND)\\

\PCR_\EXTEND & = \recv{\mathit{ch}}{(\tt{extend\_req},\mathit{nonce},\mathit{pf\_state},\mathit{v})}; \\
&\hspace{-0.5cm}\pilet \mathit{new\_st} = \extendPCR(\mathit{pf\_state},\tpmAccess,\mathit{v}) \pin \\
&\hspace{-0.5cm}\send{\mathit{ch}}{(\tt{extend\_resp},\mathit{nonce},\mathit{new\_st})}\\

\TPM_\UNSEAL &=\recv{\tt{os}}{\mathit{pf\_state}}; \\
&\hspace{-0.5cm}\Ifthen{\getLOCK(\mathit{pf\_state})}{\true} \\ 
&\hspace{-0.5cm}\;\;\;\; \pilet \mathit{ch} = \tpmch(\getPP(\mathit{pf\_state})) \pin \UNSEAL\\
& \hspace{-0.5cm}\tt{else }  \pilet \mathit{ch} = \tt{os} \pin \UNSEAL\\

\UNSEAL & = \recv{\mathit{ch}}{(\tt{tag\_unseal}, \mathit{blob})}; \\
&\hspace{-0.5cm}\pilet \mathit{v} = \unseal(\mathit{blob}, \getPCR(\mathit{pf\_state})) \pin  \\
&\hspace{-0.5cm}\send{\mathit{ch}}{(\tt{tag\_plain}, \mathit{v}))}\\ 
\hline
\end{array}
\]
\caption{The TPM process}\label{fig-TPM}
\end{figure}

We model the \TPM\ by the process in Fig.~\ref{fig-TPM}. A \PCR\ reset request can come either from the \CPU, and then the \PCR\ is reset to the value $\pd$ marking a dynamic root of trust, or else from the operating system. A \PCR\ extend request can come from the \CPU, from the operating system or from a private channel that the \CPU\ can establish between the \TPM\ and some other process. To unseal a value, the \TPM\ relies on the value of the \PCR\ registers recorded in the platform state that is associated to an unseal request. The corresponding equation for $\unseal$ ensures that    
this operation will succeed only if the \PCR\ values from the state match the \PCR\ values against which plain data was sealed. If a \DRT\ is running, we perform the unseal for the protected program \drtpp, on the private channel $\tpmch(\getPP(pf\_state))$; otherwise, the unsealed value is made public on channel $\tt{os}$.

\newpage

\subsection{Dynamic root of trust: launch}%
%

\begin{figure}
\[
\begin{array}{|l|}
\hline
\CPU= \;\;\;\;!\;\;\;\;\;\;\text{\textbf{(*** The \CPU\ process ***)}}\\
\text{\textbf{(* Step 1: receive a \DRT\ request *)}}\\
\;\recv{\tt{os}}{(\tt{drt\_req}, \mathit{init}, \mathit{pp}, \mathit{pf\_state})}\\
\;\piIfthen{\getLOCK(\mathit{pf\_state})}{\false}  \\
\;\pilet s_0'=\setINT(\mathit{pf\_state},\cpuAccess, \false) \pin\\
\;\pilet s_0 = \setLOCK(s_0', \cpuAccess, \true) \pin \\ 
\\
\text{\textbf{(* Step 2: measure \drtinit\ and the \STM\ *)}}\\
\;\pilet \mathit{measure} = (\h(\mathit{init}),\h(\getSTM(\mathit{pf\_state}))) \pin \\ 
\\
\;\text{\textbf{(* Step 3: reset and extend the \PCR\ *)}}\\
\;\pvnew{\nonce};\send{\cputpm}{(\tt{reset\_req}, \nonce, s_0)};\\
\;\recv{\cputpm}{(\tt{reset\_resp}, \nonce, s_1)}; \\
\;\send{\cputpm}{(\tt{extend\_req}, \nonce, s_1, \mathit{measure})}; \\
\;\recv{\cputpm}{(\tt{extend\_resp}, \nonce,s_2)};\\ 
\text{\textbf{(*Step 4a: load \drtinit\ \& grant \TPM\ access*)}}\\
\;\pilet s_3 = \setINIT(s_2,\cpuAccess, \mathit{init}) \pin\\
\;\pilet \mathit{einit} = \getentry(\mathit{init}) \pin \\
\;\send{\mathit{einit}}{(\nonce, s_3, \tpmch(\mathit{init}), \mathit{pp})}; \\
\;\send{\cputpm}{(\tt{ext\_channel}, \tpmch(\mathit{init})))};\\
\text{\textbf{(* Step 7b: establish \TPM\ access for \drtpp\ *)}}\\
\;\recv{\mathit{einit}}{(\tt{drt\_resp}, \nonce, \mathit{new\_state})}; \\ 
\;\pilet \mathit{epp} = \getentry(\getPP(\mathit{new\_state})) \pin \\
\;\send{\mathit{epp}}{(\mathit{new\_state}, \tpmch(\program(\mathit{epp})))};\\
\;\send{\cputpm}{(\tt{ext\_channel}, \tpmch(\program(\mathit{epp})))}\\
\\
\INIT = \;\;\;\text{\textbf{(*** A trusted \INIT\ program ***)}}\\ 
\;\send{\tt{os}}{\program(\Tinit)}; \send{\tt{os}}{\program(\Tstm)}; \\
\text{\textbf{(* Step 4b: receive \drtpp\ and \TPM\ channel *)}}\\
\;\recv{\Tinit}{(\nonce, \mathit{pf\_st}, \mathit{tpmc}, \mathit{pp})}; \\
\;\text{\textbf{(* Steps 5-6: extend $h(\drtpp)$ into \PCR\ *)}}\\
\;\pilet \mathit{measure} = \h(\mathit{pp}) \pin \pvnew{\tt{nonce}_1}; \\
\;\send{\mathit{tpmc}}{(\tt{extend\_req},\tt{nonce}_1, \mathit{pf\_st}, \mathit{measure})}; \\
\;\recv{\mathit{tpmc}}{(\tt{extend\_resp},\tt{nonce}_1, \mathit{ext\_st})}; \\
\text{\textbf{(* Step 7a: load \drtpp\ on platform state *)}}\\
\;\pilet \mathit{new\_st} = \setPP(\mathit{ext\_st},\Tinit, \mathit{pp}) \pin\\
\;\send{\tt{exp\_init}}{(\tt{drt\_resp}, \nonce, \mathit{new\_st}))};\\
\;\send{\tt{os}}{\mathit{new\_st}}\\ 
\hline
\end{array}
\]
\caption{\DRT\ process for \CPU\ and \INIT}\label{fig-DRT-CPU-INIT}
\end{figure}

The procedure for launching a dynamic root of trust, i.e. steps 1-7 from Fig.~\ref{fig:drt}, is modeled by the processes $\CPU$ and $\INIT$, from Fig.~\ref{fig-DRT-CPU-INIT}.
The \CPU\ receives a request including the \drtinit\ and \drtpp\ programs and the platform state where the \DRT\ is to be launched. If a \DRT\ is not already running in the corresponding platform state, then the \CPU\ disables the interrupts and sets the \DRT\ lock (step 1). Next, the \CPU\ measures the \INIT\ and \STM\ programs and extends the result into the \PCR\ (steps 2-3). In step 4a, the \INIT\ program is loaded and we use the term $\tpmch(\tt{init})$ to model an established private channel between the \TPM\ and the running \drtinit\ program. We use the program abstraction introduced in section~\ref{sec:prog-model} to model the loading and the execution of \drtinit, relying on the private constant \Tinit. In turn, the loaded \drtinit\ program measures the \drtpp\ program, records the measurement into the \TPM, and loads \PP\ on the platform state (steps 4b-7a). After the \drtinit\ program has measured the \drtpp\ program and loaded it into memory, the \CPU\ gets back the new platform state and sets up the private channel for communication between the loaded \drtpp\ and the \TPM\ (step 7b).

\subsection{Dynamic root of trust: execution}

\begin{figure}
\[
\begin{array}{|l|}
\hline
\PP = \textbf{(* Example of protected program *)}\\
\text{\textbf{(* Step 7c: launch and get \TPM\ access *)}}\hspace{-0.3cm}\\
\send{\tt{os}}{\program(\Tpp))};\\
\recv{\Tpp}{(\mathit{pf\_state_0},\mathit{tpmc})}; \\
\text{\textbf{(* Re-enable interrupts *)}}\\
\pilet \mathit{pf\_st} = \setINT(\mathit{pf\_state_{0}},\Tpp,\true) \\
\pin \send{\tt{os}}{\mathit{pf\_st}};\\
\\
\text{\textbf{(* Step 8: unseal and decrypt *)}}\\
\recv{\tt{os}}{x_\seal}; \recv{\tt{os}}{x_{\enc}};\\
\send{\mathit{tpmc}}{(\tt{tag\_unseal},x_\seal)};\\
\recv{\mathit{tpmc}}{(\tt{tag\_plain},x_k)}; \\
\pilet \mathit{mess} = \sdec(x_\enc,x_k) \pin \send{\tt{os}}{\mathit{mess}};\\
\\
\text{\textbf{(* Step 9: Ending the execution *)}}\\
\pvnew{\tt{rand}}; \send{\mathit{tpmc}}{(\tt{extend\_req}, \tt{rand}, \mathit{pf\_st}, \bot)}; \\
\recv{\mathit{tpmc}}{(\tt{extend\_resp}, \tt{rand}, \mathit{exts})};\\
\pilet \mathit{ends} = \setLOCK(\mathit{exts},\Tpp,\false) \pin\\
\send{\tt{os}}{\mathit{ends}}\\
\hline
\end{array}
\]
\caption{\DRT\ execution}\label{fig-DRT-execution}
\end{figure}

We illustrate the execution of a trusted \drtpp\ program with an example in Fig.~\ref{fig-DRT-execution}, where step 8 is an example of some useful execution of \drtpp, i.e., unsealing and decrypting, whereas the rest is behaviour we expect from any protected program. The private constant $\Tpp$ represents the private entry point of \PP\ according to the model from section \ref{sec:prog-model}.


In Fig.~\ref{fig-DRT-full} we consider a fresh symmetric key $k_\pp$ and assume that this key has been sealed against the measurement of the trusted \drtpp\ program, with identity $\prog(\Tinit)$, of the trusted \drtinit\ program, with identity $\prog(\Tinit)$, and of the trusted \STM\ program, with identity $\prog(\Tstm)$. This is represented by the term $\tt{sealed\_key}$ in the process $\DATA$ (see the code in the figure below), which we publish on the channel $\tt{os}$. We also assume that some private message $\hello_\pp$ is encrypted with $k_\pp$ and $\senc(\hello_\pp,k_\pp)$ is made publicly available on channel $\tt{os}$. 

In the context of a \DRT, the program $\PP$ should be able to unseal the key $k_\pp$, decrypt and publish $\hello_\pp$. Before the execution of $\PP$ ends, the \DRT\ lock is set to false, and also the \PCR\ is extended with a dummy value in order to leave the \PCR\ in a state which is not to be trusted any more. 
We verify, in section \ref{sec:model-properties}, that secret \DATA\ sealed for this program remains secret.

\begin{figure}[t]
%
\[
\begin{array}{|l|}
\hline
\DATA = \;\;\textbf{(* Seal and encrypt private data *)} \\
\pvnew{k_\pp};\pvnew{\hello_\pp};\send{\tt{os}}{\senc(\hello_\pp,k_\pp)};\\
\tt{let }\tt{sealed\_key} = \seal(k_\pp, \tt{hchain})\; \tt{in}\;\send{\tt{os}}{\tt{sealed\_key}};\\
\textbf{(* where }\tt{hchain}=\h(\h(\pd, (\h(\program(\Tinit)),\h(\program(\Tstm)))),\h(\program(\Tpp))) \textbf{ *)}\\
\\
\SETUP = \;\;\textbf{(* Launching the system *)}\\
\text{\textbf{(* Load the initial state *)}}\\
\recv{\tt{os}}{x_\stm};\recv{\tt{os}}{x_\smih};\\
\tt{out}(\tt{os},\state(\tpm(\ps),\cpu(\true,\bot),\smram(x_\stm,x_\smih), \drt(\bot,\bot,\false)); \\
\text{\textbf{(* Run a \DRT\ with any loaded programs *)}}\\
\recv{\tt{os}}{\tt{init}}; \recv{\tt{os}}{\tt{pp}}; \recv{\tt{os}}{\pfstate}; \send{\tt{os}}{(\tt{drt\_req}, \tt{init}, \tt{pp}, \pfstate)};\\

\hline
\text{\textbf{(* The main processes put together *)}}\\
\EXEC = \;\;(\; \CPU\; |\; !\;\INIT\; |\; \SETUP \; | \; \DATA \;|\; !\;\PP\;) 
\;\;\;\;\;\;\;\;\;\;\;\;\DRT = \;\;(\; \TPM\; |\; \EXEC\;) \\
\hline
\end{array} 
\]
\caption{\DRT\ setup and full process.}\label{fig-DRT-full}
\end{figure}

The $\SETUP$ process ties everything together, i.e., it loads and publishes an initial state, and runs any \DRT\ request from the operating system. We call \EXEC, all the processes put together, whereas the \TPM\ is the one providing the trusted functionalities of reset, extend, and unseal. We use $\DRT = ( \TPM\; |\; \EXEC)$.


\subsection{Security properties in the formal model}\label{sec:model-properties}


\noindent\textbf{Reachability.} The reachability of a state in the platform can be expressed as a (non-)secrecy property: a state is reachable when a corresponding state term can be obtained by the attacker after interacting with the process $\DRT$ modulo the theory $\E_\TC= \E_\fdata\cup \E_\prog \cup \E_{\state}$, expressed as a formula of the form 
\[
\DRT
\models_{\E_\TC} \Att(\state(T_\tpm,T_\cpu,T_\smram,T_\drt)) .
\]
The property that the $\DRT = ( \TPM\; |\; \EXEC)$ process can reach an expected state where some trusted programs \INIT\ and \PP\ have been correctly measured and loaded on the platform can be expressed as follows: 
\[
\begin{array}{lcl}
\begin{array}{rll}
\DRT
& \models_{\E_\TC}& \Att( \state(\\
&& \tpm(\h((\h((\pd,v_1)),v_2))),\cpu(\true,x),\\
&& \smram(\program(\Tstm),\program(y))\\
&& \drt(\program(\Tinit),\program(\Tpp),\true)) )
\end{array}	

&&
\begin{array}{rcl}
where\\
v_1 & = & (\h(\program(\Tinit)),\\
 && \;\;\h(\program(\Tstm))\\
v_2 & = & \h(\program(\Tpp)).
\end{array}
\end{array}
\]

An additional reachability property of interest is whether the program $\PP$ has succeeded to unseal the key $k_\pp$, decrypt the private message $\hello_\pp$ and output it on the public channel $\tt{os}$. This is captured by the following (non-)secrecy formula:  
\[
\DRT
\models_{\E_\TC} \Att( \hello_\pp ).
\]


\noindent\textbf{Code integrity.} We say that the trusted platform ensures code integrity if the measurement contained in the \PCR\ value correctly reflects the state of the platform. Specifically, we require that whenever a dynamic root of trust is active with a \PCR\ value of $\pd$ extended with the expected measurements $v_1$ and $v_2$, then only the corresponding \drtpp, \drtinit\ and \STM\ are running on the platform, and they cannot be modified. This can be expressed by the following correspondence assertion, which we will \emph{denote by $\Phi_\int$} in the rest of the paper: 
\[
\begin{array}{rll}
\DRT
& \models_{\E_{\TC}} & \Att( \state( \tpm(\h((\h((\pd,v_1)),v_2))),\cpu(x,y),\smram(x_\stm,x_\smih),\\
&&\;\;\;\;\;\; \drt(x_\init,x_\pp,\true)) ) \;\; \implies \;\;  (x_\init,x_\pp,x_\stm)  = (p_1,p_2,p_3)
\end{array}
\]
where $p_1  =  \program(\Tinit)$, $p_2  =  \program(\Tpp)$, $p_3  =  \program(\Tstm)$. 

Note that we ensure the property only for trusted programs. Indeed, if any of \drtpp, \drtinit\ or \STM\ are malicious, they could use their privileges to reach a platform state that does not reflect the \PCR\ values. This is fine, because the \PCR\ values will correctly record the identity of running programs in the chain of trust. In particular, our property shows that untrusted \DRT\ programs cannot make the \PCR\ values record the measurement of trusted programs. 

\noindent\textbf{Secrecy of sealed data.} We also verify that data sealed for $\PP$, i.e. the key $k_\pp$, remains secret (we \emph{denote this formula by $\Phi_\sec$}):  
\[
(\Phi_\sec)\;\;\;\;\; \DRT \models_{\E_\TC} \Att(k_\pp) \implies \mathit{false}.
\]

\section{Process transformation for automated verification} \label{sec:pcr-abstraction}

ProVerif does not terminate for the \DRT\ process and the equational theory $\E_\TC$.
The main reason is the rewrite rule from $\E_{state}$ that allows an unbounded number of \PCR\ extensions, reflecting a problem first noticed in \cite{dkrs-tpm}. In this section, we propose a general transformation of processes that allows a more efficient exploration of the search space by ProVerif. The transformation is based on a general observation formalised in Proposition \ref{prop:abs}: we can replace a process $P$ with a process $Q$ as input for ProVerif, as long as $Q$ and $P$ are equivalent with respect to the security properties of interest. Concretely, we will replace the process \DRT\ with a process $\DRT^b$ that bounds the number of \PCR\ extensions, while allowing a direct way for the attacker to set the \PCR\ to any value that is bigger than the considered bound. 

For a process $P$, let $\Att(P)=\{ T\;|\; P\models \Att(T)\}$ be the set of terms that can be obtained by the attacker when interacting with $P$. For a set of terms $\M$, we let $\Att(\M)=\{T\;|\;\M\vdash T\}$. We notice the following.

\begin{proposition}\label{prop:abs} Let $P,Q$ be processes and $\Att(T)\implies \Phi$ be a correspondence assertion such that, for any substitution $\sigma$,  
\[T\sigma \in \Att(P)\smallsetminus \Att(Q) \implies \Phi\sigma \;\;\text{\ \ and\ \ }\;\;T\sigma \in \Att(Q)\smallsetminus \Att(P) \implies \Phi\sigma.\]
Then we have: \hspace{3ex}$P\models \Att(T)\implies \Phi \text{\;\;if and only if\;\;} Q\models \Att(T)\implies \Phi$.
\end{proposition}

%

The proof of Proposition \ref{prop:abs} follows immediately from definitions, yet this result is crucial to make our models amenable for ProVerif. We are thus allowed to transform the process \DRT\ into a process $\DRT^b$, that is equivalent to \DRT\ with respect to code integrity and secrecy properties $\Phi_\int$ and $\Phi_\sec$, and whose search space can be handled by ProVerif. It will be easier to express $\DRT^b$ using some additional rewrite rules. In conjunction with Proposition \ref{prop:abs}, we will then rely on the following result for soundness and completeness: 

\begin{proposition}\label{prop:eqs} Let $\P$ be a process, $\E$ be an equational theory and $\Att(T)\implies\Phi$ be a correspondence assertion. Assume $\E^b$ is a set of rewrite rules such that $\forall U\rto V\in \E^b : \top(U)\in \F_\priv$, i.e., is a private symbol. Then we have: 
\[P\models_\E \Att(T)\implies \Phi \text{\;\;if and only if\;\;} P\models_{\E\cup \E^b} \Att(T)\implies \Phi.\]
\end{proposition}


\emph{Notation.} We denoted a term of the form $h((\ldots h((T_0,T_1))),\ldots, T_n))$ by $\chain(T_0,\ldots,T_n)$, using $\chain(T_0)$ for $T_0$. We define $\length(\chain(T_0,\ldots,T_n))=n$, representing the number of extensions of a \PCR. 

\emph{Problematic rewrite rule.} We recall the rewrite rule that poses non-termination problems for ProVerif: 

\centerline{$
\extendPCR(\state(\tpm(y),x_1,x_2,x_3), \tpmAccess, v) \rto \state(\tpm(\h((y,v))),x_1,x_2,x_3)\\
$}

\noindent
Intuitively, ProVerif does not terminate because it is unable to make an abstract reasoning about the introduction of the term $\h((y,v))$ in the right hand side of this rewrite rule. We propose a transformation of the \TPM\ process into a process $\TPM^b$ that allows more values to be written into the \PCR, overapproximating the effect of the problematic rewrite rule. This transformation will be sound and complete (satisfying the conditions of Proposition \ref{prop:abs}) based on the observation that, once it exceeds a certain bound, the value of the \PCR\ does not matter for $\Phi_\sec$ and $\Phi_\int$ -- thus, we can let the attacker have complete control over it.  

\emph{Proposed transformation.} For a given natural number $b$, we would like the following behaviour of the $\TPM^b$ process: if an extend request is received for a platform state $\state(\tpm(T_1),T_2,T_3,T_4)$ and a value $V$:

\noindent $\bullet$ if the length of the \PCR\ is smaller than $b$, i.e. $\length(T_1)<b$, then execute this request normally, using the function $\extendPCR$. The updated platform state returned by the $\TPM^b$ should now be  $\state(\tpm(h((T_1,V))),T_2,T_3,T_4)$. 
 
\noindent $\bullet$ if the length of the \PCR\ value $T_1$ is greater or equal to $b$, i.e. $\length(T_1)\geq b$, then output $T_1$ and $V$ to the attacker and wait for a new value $T_1'$ as a response. If the length of $T_1'$ is big enough, i.e. $\length(T_1')>b$, the updated platform state returned by the $\TPM^b$ should now be  $\state(\tpm(T_1'),T_2,T_3,T_4)$. In a normal execution, we would have $T_1'=h((T_1,V))$. However, the attacker has the choice to set $T_1'$ to any value.

Formally, the $\TPM^b$ process relies on the private function $\isSMALL$ to detect if the value of the $\PCR$ is lower or higher than the bound, and treat the two cases differently. The following set of rewrite rules, for all $0\leq i<b$, define $\isSMALL$: 
\hspace{3ex}$\isSMALL(\chain(v_0,\ldots,v_i))\rto\true$, \hspace{2ex}
where $v_0\in\{\ps,\pd\}$ and $v_1,\ldots,v_i$ are mutually distinct variables. We also need to check if some value to be extended into the \PCR\ is big enough. For this, we introduce the private function $\isBIG$, together with the rewrite rule: 
\hspace{3ex}$\isBIG(\chain(v_0,\ldots,v_{b+1}))\rto\true$, \hspace{2ex}
where $v_0,\ldots,v_{b+1}$ are mutually distinct variables.

\begin{figure*}
\[\begin{array}{|lcl|} 
\hline
\DRT^b & = &\TPM^b\; |\; \EXEC \\
\TPM^b & = &\TPM\;\{\;\PCR_\EXTEND\mapsto \PCR_\EXTEND^b\;\}\\

\PCR_\EXTEND^b & = &\recv{\tt{ch}}{(=\tt{extend\_req},nonce,pf\_state,val)}; \\
& &\pilet pcr = \getPCR(pf\_state) \pin \\
& &\Ifthen{\isSMALL(pcr)}{\true}\;\; \PCR_\EXTEND^\SMALL \;\;\tt{else}\;\;\PCR_\EXTEND^\BIG\\

\PCR_\EXTEND^\SMALL & = &\pilet new\_st = \extendPCR(pf\_state,\tpmAccess,val) \pin\\
& & \send{\tt{ch}}{(extend\_resp,nonce,new\_st)}\\

\PCR_\EXTEND^\BIG & = &\send{\tt{os}}{(pcr,val)}; \recv{\tt{os}}{new\_pcr}\\
 & & \Ifthen{\isBIG(new\_pcr)}{\true} \\ 
 & & \pilet new\_st = \setPCR(pf\_state,\tpmAccess,new\_pcr) \pin \\
 & &\send{\tt{ch}}{(extend\_resp,\nonce,new\_st)}\\
\hline
\end{array}
\]
\end{figure*}

The only difference from the normal \TPM\ process is in $\PCR_\EXTEND^b$, which first detects if the current value of the \PCR\ is small or big: if it is small, the extension process proceeds normally (the process $\TPM_\EXTEND^\SMALL$); if it is bigger than the given bound, then the \TPM\ requests that the operating system combines $\tt{pcr}$ and $\tt{val}$ itself (the process $\TPM_\EXTEND^\BIG$). Upon receiving the response from the $\tt{os}$, the \TPM\ first checks that the value provided is indeed big (the compromised operating system may be cheating). Only then, it updates the \PCR\ to the requested value. 

We denote by $\E_\TC^b$ the equational theory $\E_\TC$ augmented with the rules for $\isSMALL$,$\isBIG$ and $\setPCR$ introduced in this section and we assume that these new symbols are private (they are used only by $\TPM^b$).

\subsection{Sketch of correctness proofs}\label{sec:formal} 

We have to show that, for $\Phi\in \{\Phi_\sec,\Phi_\int\}$, we have $\DRT \models_{\E_\TC}\Phi \Leftrightarrow \DRT^b \models_{\E^b_\TC}\Phi$. We note that soundness (direction $\Leftarrow$) is the property that is necessary to derive the security guarantees for $\DRT$, while completeness is secondary: it explains why we dont get false attacks against $\DRT^b$ with ProVerif. Since $\Att(\DRT)\subseteq \Att(\DRT^b)$, soundness is easy to prove, while completeness requires careful analysis of terms in $\Att(\DRT^b)\smallsetminus \Att(\DRT)$. We show that such terms are roughly limited to what we explicitly release in $\DRT^b$: state terms with big \PCR\ values; they cannot be used by the attacker to violate $\Phi_\sec$ and $\Phi_\int$.  

First, from Proposition \ref{prop:eqs} and the definition of $\E_\TC^b$, we can easily translate between $\E_\TC$ and $\E_\TC^b$, thus the notions and results that follow are modulo $\E_\TC^b$.  

\begin{corollary}\label{cor:reduc-eq} For any $\Phi$, we have $\DRT\models_{\E_\TC}\Phi \Leftrightarrow \DRT\models_{\E_\TC^b}\Phi$.
\end{corollary}


Terms $T$ with $\top(T)\!=\!\state$ are called state terms (or states). For a state term $T\!=\!\state(\tpm(T_1),\cpu(T_2,T_3),\smram(T_3,T_4),\drt(T_5,T_6,T_7))$, we let $\Comp(T)\!=\!\{T_1,\dots,T_7\}$. 
For a set of terms $\M_1$, we say that a set of state terms $\M_2$ is $\M_1$-saturated if for any $T\in \M_2$ we have $\forall U\in \Comp(T):\M_1\vdash U$.  
\begin{lemma}\label{lemma:big-pcr} Let $\M_1$ be a set of terms and $\M_2$ be an $\M_1$-saturated set of state terms. Then we have $\Att(\M_1\cup \M_2)=\Att(\M_1) \cup \M_2$. 
\end{lemma}


Lemma \ref{lemma:big-pcr} formalizes the intuition that, without access to \TPM\ or \CPU, the only operation that an attacker can perform on a state is to extract its components. The proof follows by a straightforward inspection of rewrite rules. 
To help in the sequel, we consider several restrictions of attacker's power against $\DRT^b$:\\
$\bullet$ $\Att_0(\DRT^b)$ is the set of terms that can be obtained by an attacker interacting with $\DRT^b$, while not being allowed to use terms in $\Att(\DRT^b)\smallsetminus \Att(\DRT)$ when constructing inputs for $\DRT^b$. That is, $\Att_0(\DRT^b)$ can be seen as a passive attacker with respect to the additional functionality in $\DRT^b$.\\ 
\noindent$\bullet$ $\Att_1(\DRT^b)$ is the knowledge of the previous attacker whose power is augmented with the ability to unseal terms from $\Att_0(\DRT^b)$, with $\TPM\_\UNSEAL$, relying on state terms from $\Att(\DRT^b)\smallsetminus \Att(\DRT)$. This attacker is not allowed to use terms from $\Att(\DRT^b)\smallsetminus \Att(\DRT)$ in any other way.  \\
\noindent$\bullet$ $\Att_2(\DRT^b)$ is the knowledge of a \emph{state respecting} attacker against $\DRT^b$: the attacker is given unrestricted access to $\DRT^b$ and can use any terms from $\Att(\DRT^b)\smallsetminus \Att(\DRT)$ to construct his inputs; however, the attacker can only use state terms according to the specification of an honest behaviour while interacting with the \TPM, the \CPU, or the equational theory. 

Note that $\Att_0(\DRT^b)\subseteq \Att_1(\DRT^b)\subseteq \Att_2(\DRT^b) \subseteq \Att(\DRT^b)$. We denote by $\M^b$ the set of state terms returned to the attacker by the $\PCR_\EXTEND^\BIG$ process. Note that $\M^b$ is an $\Att(\DRT)$-saturated set of state terms with $\forall T\in \M^b:\length(\getPCR(T))>b$.


\begin{lemma}\label{lemma:attb} For any $b$, we have $\Att(\DRT)\subseteq \Att_0(\DRT^b)\subseteq \Att(\DRT)\cup \M^b$.
\end{lemma}
The first inclusion follows easily from the definition of $\DRT^b$, which is able to simulate any normal $\PCR$ extension performed by $\DRT$, without access to any terms in $\Att(\DRT^b)\smallsetminus \Att(\DRT)$. For the second inclusion, relying on the fact that $\M^b$ is $\Att(\DRT)$-saturated, we use Lemma \ref{lemma:big-pcr} to deduce $\Att_0(\DRT^b)\subseteq \Att(\Att(\DRT)\cup \M^b)\subseteq \Att(\DRT)\cup \M^b$. 

\begin{lemma}\label{lemma:attb2} For $b\geq 2$, we have $\Att_1(\DRT^b) \subseteq \Att_0(\DRT^b)$. 
\end{lemma}
By definition, $\Att_1(\DRT^b) \smallsetminus \Att_0(\DRT^b)\subseteq \{U\;|\; \seal(U,V)\in \Att_0(\DRT^b)\}$. Note that the only sealed term in $\Att_0(\DRT^b)$ that does not originate from the attacker is $\seal(k_\pp,\tt{hchain})$, with $\length(\tt{hchain})=2$. For any other term $\seal(U,V)\in \Att_0(\DRT^b)$, we have $U\in \Att_0(\DRT^b)$, and therefore $U\notin \Att_1(\DRT^b) \smallsetminus \Att_0(\DRT^b)$. From lemma \ref{lemma:attb}, the definition of $\TPM_\UNSEAL$, and the fact that $\forall T\in \M^b:\length(\getPCR(T))>b$, we also deduce that $k_\pp\notin \Att_1(\DRT^b) \smallsetminus \Att_0(\DRT^b)$, so we can conclude $\Att_1(\DRT^b) \subseteq \Att_0(\DRT^b)$.  

\begin{lemma}\label{lemma:attb3} For $b\geq 2$, we have $\Att_2(\DRT^b)\subseteq \Att_1(\DRT^b)\cup \M^b$. 
\end{lemma}
New terms $U \in \Att_2(\DRT^b)$ come from using a state term $V\in \Att_1(\DRT^b)$ in $\TPM_\RESET$,$\TPM_\EXTEND$ or $\CPU$. From lemmas \ref{lemma:attb} and \ref{lemma:attb2}, we have either $V\in \Att(\DRT)$ or $V\in \M^b$. In both cases, we can show that $U\in\Att_1(\DRT^b)\cup \M^b$.   


\begin{corollary}\label{cor:attb} For $b\geq 2$, we have $\Att(\DRT)\subseteq \Att(\DRT^b)\subseteq \Att(\DRT)\cup \M^b\cup \M^f$, where $\M^f$ is a set of terms such that any term $T\in \M^f$ contains a state term $T'$ with $\getPCR(T')> b$. 
\end{corollary}
The set $\M^f$ represents the additional terms that a non state respecting attacker can derive from $\M^b$.
The property of $\M^f$ is due to the fact that $\E_\TC^b$ and the $\DRT^b$ process do not have effect on state terms that are used outside their intended scope. Such terms will end up as harmless subterms of attacker's knowledge.

\begin{corollary}\label{cor:cond} For $b\geq 2$, $\DRT$ and $\DRT^b$ satisfy the conditions of Proposition \ref{prop:abs} with respect to both $\Phi_\sec$ and $\Phi_\int$.
\end{corollary}
Corollary \ref{cor:attb} shows that it is sufficient to check that conditions of Proposition \ref{prop:abs} are satisfied for terms $T$ in $\M^b\cup \M^f$. For $\Phi_\sec$, this follows from the fact that such terms $T$ are either state terms, or contain state terms, and therefore the key $k_\pp$ cannot be among them. For $\Phi_\int$, this follows from the fact that those state terms have \PCR\ lengths bigger than $2$, while the precondition of $\Phi_\int$ is a state term with \PCR\ length 2.     
%
%
%
%
From Corollary \ref{cor:cond} and Proposition \ref{prop:abs}, we deduce: 
\begin{corollary}\label{cor:reduc-proc} For $\Phi\in \{\Phi_\sec,\Phi_\int\}$, we have $\DRT\models_{\E_\TC^b}\Phi \Leftrightarrow \DRT^b\models_{\E_\TC^b}\Phi$. 
\end{corollary}

From Corollaries \ref{cor:reduc-eq} and \ref{cor:reduc-proc}, we conclude:  

\begin{theorem}\label{thm:reduc} For $\Phi\in \{\Phi_\sec,\Phi_\int\}$, $\DRT\models_{\E_\TC}\Phi \Leftrightarrow \DRT^b\models_{\E_\TC^b}\Phi$. 
\end{theorem}

\section{Verification}\label{sec:verif} 

The ProVerif code for the $\DRT^b$ process and the security properties defined in sections \ref{sec:model} and \ref{sec:pcr-abstraction} is available online\footnote{\url{www.dropbox.com/s/cvq4op3w106868t/drt.pi} \hspace{3ex} (using ProVerif version 1.85).}.
It uses the equational theory $\E_\fdata\cup \E_\prog \cup \E_{\state}^b$, 
with 
$b=2$. 
The verification of each security property terminates in order of minutes, 
returning the expected result. From these results (implying there is no attack on $\DRT^b$ modulo $\E_\TC^b$) and from Theorem \ref{thm:reduc} (implying there is no attack on $\DRT$ modulo $\E_\TC$), we derive: 

\begin{theorem} The \DRT\ process satisfies, modulo $\E_\fdata\cup \E_\prog\cup \E_{\state}$, the properties of code integrity and data secrecy defined in section \ref{sec:model-properties}.
\end{theorem}


In order to check the reachability properties $\DRT\models\Phi$ defined in section \ref{sec:model-properties}, we give $\neg (\DRT\models\Phi)$ as input query for ProVerif - an attack with respect to this query would be a witness trace for the desired reachability property. When returning such a trace, ProVerif can either confirm that it is valid (\emph{attack found}) or cannot confirm it. Our models fall in the latter case, and we have to further inspect the output trace to see how its steps can be used to reconstruct a valid trace: we do observe in the output trace the expected intermediary messages on the channels $\cputpm$ and $\tt{os}$, and we can follow the source of these messages up to a dynamic root of trust request, of whose validity we have to again make sure. By a similar analysis of attack traces returned by ProVerif, we can observe the attacks of \cite{txt-att1,txt-att2} in our models, when we allow the \STM\ to be modified arbitrarily. 



%
%

\section{Further work}
%

While our model takes into account at an abstract level the attacks and mitigations of \cite{txt-att1,txt-att2}, further refinements and soundness results are necessary in order to be able to conclude that attacks such as these or as \cite{txt-att3,txt-att4} are not possible in practice. We need to develop models that are abstract enough to allow clear specifications and automated reasoning, and realistic enough to capture for instance implementation flaws. We plan to see how the models of this paper can be expressed in richer frameworks like StatVerif \cite{statverif} and SAPIC \cite{KremerK14GlobalState}, in order to capture more closely the state semantics of real platforms. 
%
%
We think the process transformation that we have presented in section \ref{sec:pcr-abstraction} is an instance of a more general result, whose exploration would also be fruitful for future applications.  

\newpage

\bibliography{drt}
\bibliographystyle{abbrv}

\longVersion{
\newpage
  \appendix

\section{Operational semantics of the process calculus} 

\begin{figure}
\[
\begin{array}{ll} 
\tt{(NIL)} & (\N,\M,\P\cup\{0\})\rto (\N,\M,\P)\\
\tt{(BANG)} &(\N,\M,\P\cup\{!P\})\rto (\N,\M,\P\cup \{P,!P\})\\
\tt{(PAR)} &(\N,\M,\P\cup\{P\;|\;Q\})\rto (\N,\M,\P\cup \{P,Q\})\\
\tt{(NEW)} &(\N,\M,\P\cup\{\pvnew{n}; P\})\rto (\N\cup \{n'\},\M,\P\cup \{P\})\\
  &\;\;\;\;\;\;\;\;\;\text{where }n'\notin \N\\
\tt{(COMM)} &(\N,\M,\P\cup\{\send{U}{T};P\;,\;\recv{U}{A};Q\})\rto (\N,\M',\P\cup \{P,Q\sigma\downarrow\})\\
  &\;\;\;\;\;\;\;\;\;\text{where }\sigma \text{ is such that }T=_\E A\sigma \text{ and}   \\
& \;\;\;\;\;\;\;\;\;\text{if }\M\vdash_\E U, \text{ then }\M'=\M\cup\{T\};\text{ else, }\M'=\M   \\
\tt{(IF}_\tt{T}\tt{)} & (\N,\M,\P\cup\{\piIfthenelsep{U = V}{P}Q\})\rto (\N,\M,\P\cup \{P\})\\
&\;\;\;\;\;\;\;\;\;\text{if }U=_\E V\\
\tt{(IF}_\tt{F}\tt{)} &(\N,\M,\P\cup\{\piIfthenelsep{U = V}{P}Q\})\rto (\N,\M,\P\cup \{Q\})\\
&\;\;\;\;\;\;\;\;\;\text{if }U\neq_\E V\\
\tt{(LET)} &(\N,\M,\P\cup\{\Letin{x}{T}{P}\})\rto (\N,\M,\P\cup \{P[x\mapsto T]\downarrow\})\\
&\;\;\;\;\;\;\;\;\;\text{if }T\downarrow\text{ does not contain destructors}\\
\end{array}
\]
\caption{Operational semantics \label{fig:semantics}}
\end{figure}

\section{ProVerif code}
\begin{verbatim}
 
(***
ABBREVIATIONS: 
DRT - DYNAMIC ROOT OF TRUST 
DRT_INIT - THE SINIT (INTEL) OR SLB (AMD) PROGRAM
DRT_PP - THE PROTECTED PROGRAM: MLE(INTEL) OR SK(AMD)
***)

param reconstructTrace = false.

(*CHANNELS*) 
free os. (* PUBLIC CHANNEL FOR THE OPERATING SYSTEM CONTROLLED BY THE INTRUDER *)	
private free cpu_tpm.  (* PRIVATE CHANNEL FOR THE COMMUNICATION BETWEEN CPU AND TPM *)
private fun tpm_ch/1. (* tpm_ch(x) REPRESENTS A PRIVATE CHANNEL USED BY 
          A PROGRAM x TO COMMUNICATE WITH TPM *)
(*CRYPTO*)
fun h/1.	(* HASH FUNCTION *)
fun senc/2.	(* SYMMETRIC ENCRYPTION *)
reduc sdec(x,senc(x,y)) = y.
fun ps/0.  (* STATIC RESET VALUE OF THE PCR *)
fun pd/0.  (* DYNAMIC RESET VALUE OF THE PCR *)
fun false/0. fun true/0. (* BOOLEAN VALUES *) 

(* TPM SEAL/UNSEAL*)      
fun seal/2. 
private reduc unseal(seal(xpcr,xvalue), xpcr) = xvalue. 

(* STATE STRUCTURE: state(tpm(PCR),cpu(INT,CACHE), drt(INIT,PP,LOCK),smram(STM,SMI) *) 
private fun state/4. private fun tpm/1. private fun cpu/2. 
private fun drt/3.  private fun smram/2. 

(*EXAMPLE STATE: 
state(tpm(pd),cpu(true,false), drt(program(expected_init),program(expected_pp),true),
smram(program(expected_stm),program(expected_smih)))
*) 


(* PRIVATE CONSTANTS FOR THE PRIVILEGED ACCESS THAT 
THE CPU AND TPM HAVE TO THE PLATFORM STATE *)
private fun cpuAccess/0. private fun tpmAccess/0. 

(* ABSTRACTION FOR DYNAMICALLY LOADING PROGRAMS*)
fun program/1. 
private reduc getENTRY(program(x)) = x. 


(*** ACCESSING THE PLATFORM STATE ***)
reduc getPCR (state(tpm(y),x1,x2,x3)) = y. 
reduc getINT(state(x1,cpu(y1,y2),x2,x3)) = y1.
reduc getCACHE(state(x1,cpu(y1,y2),x2,x3)) = y2. 
reduc getINIT(state(x1,x2,drt(y1,y2,y3),x3)) = y1.
reduc getPP(state(x1,x2,drt(y1,y2,y3),x3)) = y2.
reduc getLOCK(state(x1,x2,drt(y1,y2,y3),x3)) = y3. 
reduc getSTM (state(x1,x2,x3,smram(y1,y2))) = y1. 
reduc getSMIH (state(x1,x2,x3,smram(y1,y2))) = y2. 


(*** MODIFYING THE PLATFORM STATE + ABILITIES OF LOADED PROGRAMS ***)
(* TPM *)
reduc resetPCR (state(tpm(y),x1,x2,x3),tpmAccess,pd)		=state(tpm(pd),x1,x2,x3);
      resetPCR (state(tpm(y),x1,x2,x3),tpmAccess,ps)		=state(tpm(ps),x1,x2,x3).


reduc 
    (*** PROBLEMATIC EQUATIOn *)
      extendPCR(state(tpm(x),x1,x2,x3), tpmAccess, value) = 
      state(tpm(h((x,value))),x1,x2,x3).
    (***)
    
reduc setPCR(state(tpm(y),x1,x2,x3),tpmAccess,value)		=state(tpm(value),x1,x2,x3).

reduc isSMALL(pd) = true; 
      isSMALL(ps) = true; 
      isSMALL(h((pd,y))) = true; 
      isSMALL(h((ps,y))) = true. 
      
reduc isBIG(h((h((h((x,y)),z)),w))) = true.       
      

(* CPU *) 
reduc setINT(state(x1,cpu(y1,y2),x2,x3),cpuAccess,value) = state(x1,cpu(value,y2),x2,x3);
      setINT(state(x1,cpu(y1,y2),drt(z1,program(z2),true),x2),z2,value) = 
      state(x1,cpu(value,y2),drt(z1,program(z2),true),x2).
reduc cache(state(x1,cpu(y1,y2),x2,x3),value) = state(x1,cpu(y1,value),x2,x3).
reduc flush_smi(state(x1,cpu(y1,y2),x2,smram(z1,z2))) = 
      state(x1,cpu(y1,y2),x2,smram(z1,y2)). 
reduc flush_stm(state(x1,cpu(y1,y2),drt(w1,w2,false),smram(z1,z2))) = 
      state(x1,cpu(y1,y2),drt(w1,w2,false),smram(y2,z2)).
(* TO OBTAIN THE ATTACK, ADD THE EQUATION: *)
(* flush_stm(state(x1,cpu(y1,y2),x2,smram(z1,z2))) = state(x1,cpu(y1,y2),x2,smram(y2,z2)). *)

(* DRT *)
reduc setINIT(state(x1,x2,drt(y1,y2,y3),x3),cpuAccess,value) 	=
      state(x1,x2,drt(value,y2,y3),x3).
reduc setPP(state(x1,x2,drt(y1,y2,y3),x3),cpuAccess,value)	=
      state(x1,x2,drt(y1,value,y3),x3);
      setPP(state(x1,x2,drt(program(y1),y2,y3),x3),y1,value)	=
      state(x1,x2,drt(program(y1),value,y3),x3);
      setPP(state(x1,cpu(true,z),drt(y1,y2,y3),
            smram(program(z1),program(z2))),(z1,z2),value)=
      state(x1,cpu(true,z),drt(y1,value,y3),smram(program(z1),program(z2))).
reduc setLOCK(state(x1,x2,drt(y1,y2,y3),x3),cpuAccess,value)	=
      state(x1,x2,drt(y1,y2,value),x3);
      setLOCK(state(x1,x2,drt(y1,program(y2),true),x3),y2,value)=
      state(x1,x2,drt(y1,program(y2),value),x3);
      setLOCK(state(x, cpu(true,z),drt(y1,y2,y3),
      smram(program(z1),program(z2))),(z1,z2),value)
	=state(x,cpu(true,z),drt(y1,y2,value),smram(program(z1),program(z2))).
							
      
(** MESSAGE TAGS **)		
free drt_request,drt_response,pcr_extend_request,pcr_extend_response,
pcr_reset_request,pcr_reset_response, drt_start, tag_unseal, tag_plain, ext_channel.
 
(* FUNCTION FOR CREATING NONCES *) 
private fun fnonce/1. 

let DRT_CPU = (* GET A DRT REQUEST FROM THE OPERATING SYSTEM *)
  		  in(os, (=drt_request, drt_init, drt_pp, pf_state));
		  (* ONLY ACCEPT THE REQUEST IF NOT ALREADY RUNNING A DYNAMIC ROOT OF TRUST *)
		  if getLOCK(pf_state) = false then 
		  (
		   (* DISABLE INTERRUPTS *) 
		   let s0'=setINT(pf_state,cpuAccess, false) in
		   (* UPDATE THE LOCK *) 
		   let s0 = setLOCK(s0', cpuAccess, true) in 

		   (* RESET THE PCR *)
		   (* DESIRED CODE: *)
		   (* new nonce; *)
		   (* CLASSIC ABSTRACTION THAT RUNS FASTER: NONCES ARE A FUNCTION OF THEIR CONTEXT *)
		   let nonce = fnonce((drt_init,drt_pp, getSTM(pf_state))) in  
		    
		   out(cpu_tpm, (pcr_reset_request, nonce, s0)); 
		   in(cpu_tpm, (=pcr_reset_response,=nonce,s1));
		    
		   (* EXTEND THE PCR WITH THE MEASUREMENT *)
		   let measurement = (h(drt_init),h(getSTM(pf_state))) in 
		   out(cpu_tpm, (pcr_extend_request, nonce, s1, measurement)); 
		   in(cpu_tpm, (=pcr_extend_response, =nonce,s2));
		    
		   (* LOAD DRT_INIT AND ESTABLISH TPM CHANNELS *)
		   let s3 = setINIT(s2,cpuAccess, drt_init) in
		   out(cpu_tpm, (ext_channel, tpm_ch(drt_init)));
		   let entry_init = getENTRY(drt_init) in 
		   out(entry_init, (drt_request, nonce, s3, tpm_ch(drt_init), drt_pp)); 
		    
		   (* THE drt_init PROGRAM HAS MEASURED AND SET UP THE drt_pp PROGRAM*) 
		    in(entry_init, (=drt_response, =nonce, new_state)); 
		    
		   (* SETUP TPM CHANNELS FOR THE LOADED DRT_PP *) 
		    let entry_pp = getENTRY(getPP(new_state)) in 
		    out(entry_pp, (drt_start, new_state, tpm_ch(program(entry_pp))));
		    out(cpu_tpm, (ext_channel, tpm_ch(program(entry_pp))))
		    
		   ). 

(* 
THE TWO EQUATIONS BELOW TOGETHER WITH THE CACHE PROCESS ARE A CONSEQUENCE OF
cache, flush_smi,flush_stm EQUATIONS. WRITING THEM EXPLICITLY HELPS PROVERIF 
TERMINATE 5 MINUTES FASTER 
*)
reduc setSTM (state(x1,x2,x3,smram(y1,y2)),cpuAccess,value)	=
state(x1,x2,x3,smram(value,y2)). 
reduc setSMIH(state(x1,x2,x3,smram(y1,y2)),cpuAccess,value)	=
state(x1,x2,x3,smram(y1,value)). 
let CACHE = ( in(os, (pf_state,xsmi)); 
		let new_state = setSMIH(pf_state,cpuAccess, xsmi) in 
		out(os, new_state) )
		|
		( in(os,(pf_state,xstm));
		if getLOCK(pf_state) = false then 
		let new_state = setSTM(pf_state,cpuAccess,xstm) in 
		out(os, new_state) ).
		   
private fun expected_init/0.
let EXPECTED_INIT = out(os, program(expected_init)); 
		    (* RECEIVE DRT_PP AND TPM ACCESS FROM THE CPU *)
		    in(expected_init, (=drt_request, nonce0, pf_state, tpmc, drt_pp)); 
		    (* MEASURE AND EXTEND DRT_PP INTO THE PCR *)
		    let measurement = h(drt_pp) in 
		    (* DESIRED CODE: *)
		    (* new nonce; *)
		    (* CLASSIC ABSTRACTION THAT RUNS FASTER: NONCES ARE A FUNCTION OF THEIR CONTEXT *)
		    let nonce = fnonce(drt_pp) in 
		    out(tpmc, (pcr_extend_request,nonce, pf_state, measurement)); 
		    in(tpmc, (=pcr_extend_response,=nonce, ext_state)); 
		    (* LOAD DRT_PP ON THE PLATFORM STATE *)
		    let new_state = setPP(ext_state,expected_init, drt_pp)  in
		    (* PASS THE CONTROL BACK TO THE CPU *)
		    out(expected_init, (drt_response, nonce0, new_state));
		    (* MAKE THE NEW PLATFORM STATE PUBLIC *)
		    out(os, new_state). 

private fun expected_pp/0. 
let EXPECTED_PP = (* DECRYPT A SEALED BLOB, RELYING ON COMMUNICATION WITH TPM*)
		  out(os, program(expected_pp)); 
		  in(expected_pp, (=drt_start,pf_state0,tpmc)); 
		  (* RE-ENABLE INTERRUPTS *)
		  let pf_state = setINT(pf_state0,expected_pp,true) in 
		  out(os,pf_state);
		  (* UNSEAL THE KEY AND DECRYPT THE PRIVATE MESSAGE *)		  
		  in(os,xSealedBlob);  in(os,xEncBlob);
		  out(tpmc,(tag_unseal,xSealedBlob));
		  in(tpmc,(=tag_plain,xSymKey)); 
		  let xMessage = sdec(xSymKey,xEncBlob) in
		  out(os,xMessage);

		  (* ENDING THE EXECUTION: THE LOCK IS SET TO FALSE AND THE PCR VALUE IS DESTROYED *)
		  (*new nonce; *)
		  (* ABSTRACTION THAT RUNS FASTER *)
		  let nonce = fnonce(drt_pp) in 
		  out(tpmc, (pcr_extend_request, nonce, pf_state, zero)); 
		  in(tpmc, (=pcr_extend_response, =nonce, ext_state));
		  let end_state = setLOCK(ext_state,expected_pp,false) in
		  out(os,end_state).

let TPM = !TPM_RESET | !TPM_EXTEND | !TPM_UNSEAL.  

let TPM_RESET = let (channel, reset_type) = (cpu_tpm,pd) in !PCR_RESET |
		 let (channel, reset_type) = (os,ps) in !PCR_RESET.  
		 
let PCR_RESET = in(channel, (=pcr_reset_request, nonce, pf_state));
		let new_state = resetPCR(pf_state,tpmAccess,reset_type) in  
		out(channel, (pcr_reset_response, nonce, new_state)). 

let TPM_EXTEND = let channel = os in !PCR_EXTEND_BOUND |
		 let channel = cpu_tpm in !PCR_EXTEND_BOUND | 
		 !in(cpu_tpm, (=ext_channel, channel)); !PCR_EXTEND_BOUND.

		 
let PCR_EXTEND = in(channel, (=pcr_extend_request,nonce,pf_state,value)); 
		 let new_state = extendPCR(pf_state,tpmAccess,value) in 
		 out(channel, (pcr_extend_response,nonce,new_state)).
		 
		
let PCR_EXTEND_BOUND = 
		 in(channel, (=pcr_extend_request,nonce,pf_state,value)); 
		 let pcr  = getPCR(pf_state) in 
		 if isSMALL(pcr)=true then 
		 PCR_EXTEND_SMALL else PCR_EXTEND_BIG.

let PCR_EXTEND_SMALL
	      = let new_state = extendPCR(pf_state,tpmAccess,value) in 
		out(channel, (pcr_extend_response,nonce,new_state)).


let PCR_EXTEND_BIG 
	      = 
	      out(os,(pcr,val)); in(os,new_pcr); 
              if isBIG(new_pcr)=true then  
              let new_state = setPCR(pf_state,tpmAccess,new_pcr) in 
	      out(channel, (pcr_extend_response,nonce,new_state)).
		 
let TPM_UNSEAL = in(os, (=tag_unseal, pf_state, blob)); 
	      	 let value = unseal(blob, getPCR(pf_state)) in  
		 if getLOCK(pf_state) = true then 
		  (  let channel = tpm_ch(getPP(pf_state)) in 
		     out(channel, (tag_plain, value)) )
		 else 
		     out(os, (tag_plain, value)).

		
(* QUERIES *)				
(*A. THE EXPECTED STATE HAS BEEN REACHED *)
query attacker:state(tpm(h((h((pd,(h(program(expected_init)),h(program(expected_stm))))),
	             h(program(expected_pp))))), cpu(true,x), 
		      drt(program(expected_init),program(expected_pp),true),
		      smram(program(expected_stm),program(y))). 
(* QUERY RESULT: PROVERIF RETURNS AN ATTACK TRACE THAT CAN BE INSPECTED FOR VALIDITY *)

(*B. THE PROTECTED PROGRAM SUCCESFULLY DECRYPTS THE PRIVATE 
MESSAGE USING THE SEALED KEY AND MAKES IT PUBLIC*)
query attacker:hello_pp.								
(* QUERY RESULT: PROVERIF RETURNS AN ATTACK TRACE THAT CAN BE INSPECTED FOR VALIDITY *)
		
(*C. WHENEVER THE EXPECTED PCR IS SET,  
THE PLATFORM HAS THE EXPECTED STATE *)
query attacker:state( tpm(h((h((pd,(h(program(expected_init)),
                          h(program(expected_stm))))),h(program(expected_pp))))), 
		      cpu(x,y),drt(xi,xp,true),
		      smram(xstm,xsmih)) 
		==> (xi,xp,xstm)=(program(expected_init),program(expected_pp),program(expected_stm)). 
		
(*QUERY RESULT: TRUE => THE ASSERTION IS VALID*)

(*D. THE ATTACKER DOES NOT HAVE ACCESS TO THE SEALED KEY *) 
query attacker:k_pp.				
(* QUERY RESULT: TRUE => THE ATTACKER DOES NOT HAVE k_pp *)

(* THE MAIN PROCESS *)
free null. 
private fun expected_stm/0. 
private free k_pp.	(* SECRET KEY WHICH SHOULD ONLY BE KNOWN BY THE PROTECTED PROGRAM *)
private free hello_pp.	(* PRIVATE MESSAGE ENCRYPTED WITH k_pp *)
process (* ENCRYPTED PRIVATE MESSAGE FOR PP *)
	out(os,senc(k_pp,hello_pp));
	(*ASSUME THAT THE BLOB SEALING THE SECRET KEY IS PUBLIC*)
	out(os,seal(h((h((pd,(h(program(expected_init)),h(program(expected_stm))))),h(program(expected_pp)))),k_pp));
	
	out(os, program(expected_stm));
	(* INITIAL STATE LOADED UPON A SYSTEM RESET *)
	in(os,(xInitStm,xInitSmih));
	out(os, state(tpm(ps),cpu(true,null),drt(null,null,false),smram(xInitStm,xInitSmih))); 
	(* REQUESTING A DYNAMIC ROOT OF TRUST WITH ANY LOADED PROGRAMS *)
	in(os, drt_init); in(os,drt_pp); in(os,pf_state); 
	out(os, (drt_request, drt_init, drt_pp, pf_state));
	(*EXECUTING THE DRT PROCESSES *)
        ( !DRT_CPU | !CACHE | !EXPECTED_INIT | !EXPECTED_PP | TPM)

\end{verbatim}

}
\end{document}